\documentclass[12pt]{article}
\pdfoutput=1

\usepackage{amsthm}
\usepackage{amsmath,amssymb,amsfonts}
\usepackage{graphicx,color}
\usepackage{multirow}
\usepackage{tabularx}
\usepackage{makecell}
\usepackage{booktabs}
\usepackage{wrapfig}
\usepackage{subcaption}
\usepackage{slashed}

\usepackage{enumitem}
\usepackage{braket}
\usepackage{bbm}
\usepackage{bm}
\usepackage{physics}
\usepackage{comment}
\usepackage{array}
\usepackage{cite}
\usepackage{mathtools}
\usepackage{framed}
\usepackage{diagbox}

\makeatletter
\newcommand{\thickhline}{%
    \noalign {\ifnum 0=`}\fi \hrule height 1pt
    \futurelet \reserved@a \@xhline
}
\newcolumntype{"}{@{\hskip\tabcolsep\vrule width 1pt\hskip\tabcolsep}}
\makeatother

\newcommand{\be}{\begin{eqnarray}}
\newcommand{\ee}{\end{eqnarray}}

\usepackage[normalem]{ulem}

\allowdisplaybreaks

\begin{document}

\begin{titlepage}

\begin{flushright}
{\small
TUM-HEP-1500/24\\
March 11, 2024\\
}
\end{flushright}

\vskip1cm
\begin{center}
{\Large \bf Enhancement of $p$-wave dark matter annihilation\\[0.1cm] 
by quasi-bound states}
\end{center}
\vspace{0.5cm}
\begin{center}
{\sc Martin~Beneke, 
\sc Tobias Binder, 
\sc Lorenzo De Ros, 
and Mathias Garny} 
\\[6mm]
{\it Physik Department T31,\\
James-Franck-Stra\ss e~1,
Technische Universit\"at M\"unchen,\\
D--85748 Garching, Germany}
\end{center}
\vskip1cm

\begin{abstract}
\noindent
We scrutinize the Sommerfeld enhancement in dark matter pair annihilation for $p$-wave and higher-$\ell$ partial waves. For the Yukawa potential these feature a super-resonant Breit-Wigner peak in their velocity-dependence close to Sommerfeld resonances as well as a universal scaling with velocity for all $\ell\geq 1$ that differs from the $s$-wave case.
We provide a quantum mechanical explanation for these phenomena in terms of quasi-bound states sustained by the centrifugal barrier of the partial-wave potential, and give approximate WKB expressions capturing the main effects. The impact of quasi-bound states is exemplified for wino dark matter and models with light mediators, with a focus on indirect detection signals.
We note that quasi-bound states can also explain similar peaks in the bound-state formation and self-scattering cross sections.
\end{abstract}

\end{titlepage}


\section{Introduction}
\label{sec:Introduction}

It has become evident that dark matter (DM) pair annihilation cannot always be accurately described by the Born process. In standard WIMP models with  DM mass exceeding the electroweak scale, and in dark-sector models with new light mediators, the low-velocity scattering via exchange of the mediators prior to annihilation significantly alters the annihilation rate through the Sommerfeld effect~\cite{Hisano:2003ec, Hisano:2004ds, Hisano:2006nn, ArkaniHamed:2008qn}. 

Analyzing this effect often requires solving a multi-dimensional Schr\"odinger equation numerically due to the matrix potential generated by  mediator exchange and coupled channels for multiple heavy particles  degenerate with the DM mass. For a single channel, the Sommerfeld effect is represented by a factor, $S_\ell$, in the partial-wave expanded annihilation cross section, given by
\begin{align}
(\sigma v)_\ell  = c_\ell v^{2\ell} S_\ell\, 
\label{eq:ldecomposition}
\end{align}
where $c_\ell$ are model-dependent coefficients. The combination $c_\ell v^{2\ell}$ represents the short-distance part, usually identified by the velocity-expanded tree-level annihilation cross section. The case of $s$-wave annihilation is well understood, while the higher $\ell$-wave annihilations are often regarded as numerically subdominant.
 
For a massless mediator, the induced potential is of Coulomb-type,  $V(r)=-\alpha/r$, with an effective coupling strength $\alpha$ for the underlying particle physics model. The corresponding Sommerfeld factors for different partial waves have analytic expressions~\cite{Sommerfeld:1931qaf, Sakharov:1948yq, Iengo:2009ni, Cassel:2009wt}
\begin{equation}
S_\ell = \frac{\pi/\epsilon_v}{1-e^{-\pi/\epsilon_v}} 
\prod_{b=1}^\ell \left(1+\frac{1}{4 b^2 \epsilon_v^2}\right)\;,
\end{equation}
where $v$ denotes the  relative velocity and  
\begin{align}
   \epsilon_v \equiv v/ (2\alpha) \;.
   \label{eq:epsv_def}
\end{align}
In the small velocity limit $\epsilon_v \ll 1$, the Sommerfeld factor scales as $S_\ell \propto 1/\epsilon_v^{2\ell+1}$, resulting, according to Eq.~(\ref{eq:ldecomposition}), in $(\sigma v)_\ell 
\propto 1/\epsilon_v$ independent of $\ell$. However, the 
$v^{2\ell}$ suppression in the regime $\epsilon_v \gtrsim 1$ where $S_\ell \simeq 1$, and the coefficients $c_l$ usually suppress the higher partial-wave contributions. This is why only the lowest contributing partial wave is usually included for the Sommerfeld-enhanced annihilation cross section, often providing sufficient precision for estimating relevant observables. 

The situation can, however, change if the mediator is light but massive, and the induced potential is of the Yukawa-type $V(r)= -\alpha e^{-m_\phi r}/r$, with the mass of the mediator $m_\phi$. A major difference compared to the Coulomb case is that $S_\ell$ exhibits resonant behavior for specific, $\ell$-dependent values of the ratio
\begin{align}\label{eq:epsilonphi}
\epsilon_\phi \equiv m_\phi/(\alpha m_\chi)\,.
\end{align}
This behavior is known to be closely related to the presence of zero-energy bound states in the spectrum. For example, at resonant $\epsilon_\phi$ values for $s$-wave, the Sommerfeld factor scales as $S_{\ell=0} \propto \epsilon_v^{-2}$ for $\epsilon_v \lesssim \epsilon_\phi$; a scaling stronger than in the Coulomb case. For small deviations around the exact resonant $\epsilon_\phi$ value, $S_{\ell=0}$ saturates to a constant for some small $\epsilon_v$. An analytic result for the Sommerfeld factor of the Yukawa potential is unknown, but an expression exists for the Hulth\'{e}n potential with a modified centrifugal term \cite{Cassel:2009wt}. While these analytic results for the $s$-wave qualitatively explain the resonant structure and the velocity dependence of $S_{\ell=0}$ as described, for $\ell \geq 1$ they fail to reproduce numerical computations with a Yukawa potential for such parameter regimes (see Ref.~\cite{Ding:2021zzg} for a detailed $p$-wave annihilation comparison).

The present work emerges from an exploration of the Sommerfeld 
factor for the Yukawa potential for higher partial waves, in the course of 
which we observed for all $\ell \geq 1$ unexpected deviations from the small-velocity 
scaling above and resonant behavior in the velocity dependence 
at peculiar velocities (not captured by the analytic expressions in Ref.~\cite{Cassel:2009wt}). While such spikes can be seen in several studies, not only in Sommerfeld-enhanced annihilation~\cite{Ding:2021zzg, Kamada:2023iol, Biondini:2023ksj} but also in bound-state formation~\cite{Petraki:2016cnz, Biondini:2023ksj} and self-scattering cross sections~\cite{Tulin:2012wi, Kamada:2023iol}, a deeper quantum-mechanical explanation for this behavior is still missing.

In this work we identify the existence of {\em quasi-bound states} (QBS), a term following standard literature of quantum mechanics (e.g.,~\cite[§134]{Landau:1991wop}), as the origin of the anomalous scaling and spike in the velocity dependence for annihilation. The QBS have discrete but \emph{positive} energies, which are directly related to the velocity peak position of the spike. The width of the velocity spike we associate with the QBS inverse life time. Utilizing the WKB method, we analytically determine the energy and width of the QBS, and give a quantitative explanation of the previously observed $\ell \geq 1$ resonance structures for a Yukawa potential. While we focus on annihilation, corresponding peaks in bound-state formation and self-scattering cross sections~\cite{Kamada:2023iol} can also be understood via the impact of QBS.


\begin{figure}[t]
    \centering
    \includegraphics[scale=0.9]{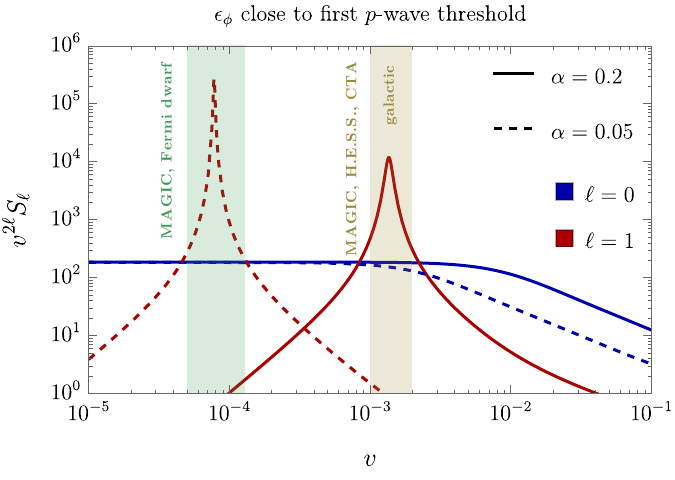}

    \caption{\small The Sommerfeld factor, $S_\ell$, for an attractive Yukawa potential can feature spikes in the velocity dependence for the $\ell \geq 1$ partial waves due to the existence of \emph{quasi-bound states} (QBS). Here, we choose for illustration the first QBS appearing in the $p$-wave (largest $\epsilon_\phi=m_\phi/(\alpha m_\chi) \approx 0.11$) and numerically compute $S_{\ell=0}$ (blue) and $S_{\ell=1}$ (red) for two different values of the effective coupling $\alpha$ (solid and dashed) entering the Yukawa potential. It is seen that the $p$-wave Sommerfeld-enhanced annihilation cross section in the presence of a QBS can dominate over the $s$-wave one by orders of magnitude. It is also demonstrated that the peak location of the DM annihilation signal, associated with the discrete positive energy of the QBS, can occur at different velocity matching the typical DM velocity in the Galactic Center and dwarf galaxies, respectively, and probed by current (Fermi Gamma-ray Space Telescope~\cite{Fermi-LAT:2015att}, H.E.S.S.~\cite{HESS:2016mib}, MAGIC~\cite{MAGIC:2022acl}) and future (Cherenkov Telescope Array~\cite{CTA:2020qlo}) indirect detection experiments.  }
    \label{fig:intro}
\end{figure}


QBS phenomena have velocity- and $\ell$-dependent characteristics, which may lead to interesting phenomenological signatures. By way of example, Fig.~\ref{fig:intro} shows the anticipated spikes in the velocity dependence of the $\ell=1$ Sommerfeld factor for the Yukawa potential as caused by the existence of the first quasi-bound state (largest $\epsilon_\phi\approx 0.11$). The numerical calculations are performed for two different coupling values of $\alpha$, demonstrating that within the range of electroweak and rather strong coupling values the peak location of the annihilation signal can occur at typical velocities on dwarf galactic and galactic scales. As the $s$-wave contribution is subdominant here, only through the QBS enhancement of the higher partial waves (here $\ell=1$) the annihilation signal may be probed by current and future indirect detection experiments. A detailed parametric dependence of the location and height of these peaks for different $\ell \geq 1$ is subject to this work.
 
The outline of this paper is as follows. In Sec.~\ref{sec:Quasi-bound_states}, we analyze the resonant structure of Sommerfeld-enhanced annihilation for the Yukawa potential at higher partial waves. Sec.~\ref{sec:WKB} provides an interpretation of the observed resonant structure in terms of QBS phenomena. 
In particular, by utilizing the WKB method we identify a factorized Breit-Wigner distribution in the wave function as a common explanation. The possible impact of quasi-bound states on cosmic-ray signals from dark matter annihilation is exemplified for DM models with light mediators and wino dark matter in Sec.~\ref{sec:Examples}. We conclude in Section~\ref{sec:Conclusion}. Technical details can be found in three appendices.


\section{Quasi-bound states}
\label{sec:Quasi-bound_states}


\subsection{Yukawa potential}
\label{sec:Yukawa_potential}

\begin{figure}[t]
    \centering
    \includegraphics[scale=0.88]{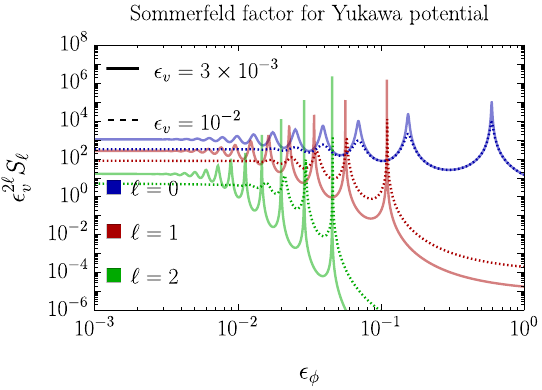}
    \includegraphics[scale=0.88]{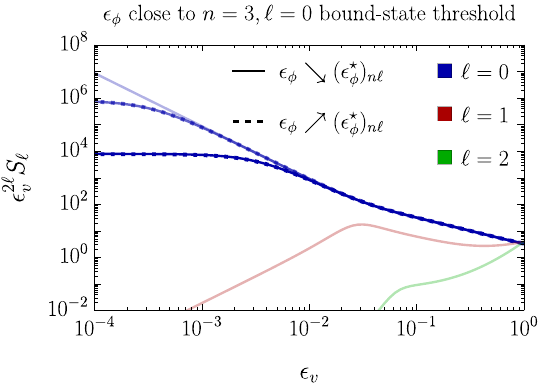}\\
    \vspace{0.5 cm}
    \includegraphics[scale=0.88]{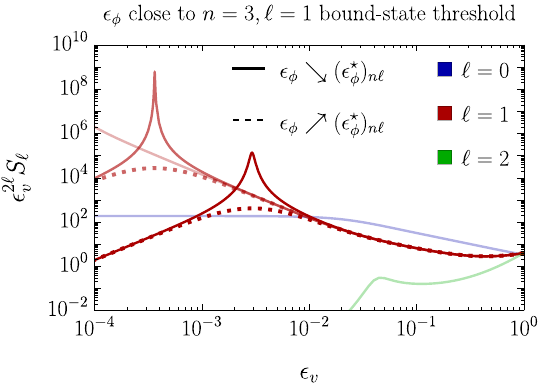}
    \includegraphics[scale=0.88]{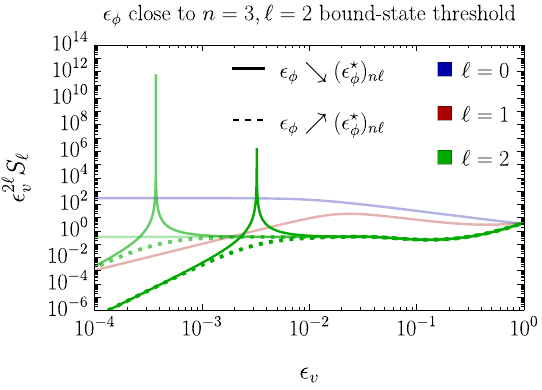}
    \caption{\small Numerical solution for the Sommerfeld factor for the attractive Yukawa potential. Note that $\epsilon_v^{2\ell} S_\ell$ is shown, which corresponds to the partial-wave annihilation cross sections, see Eq.~\eqref{eq:ldecomposition}. \emph{Upper-left panel:} Resonant structure of the three lowest partial waves as a function of $\epsilon_\phi$ for two exemplary values of $\epsilon_v$. Observe that the peak positions for $\ell \geq 1$ are velocity ($\epsilon_v$) dependent. \emph{Upper-right and bottom panels:} Velocity dependence of $\epsilon_v^{2\ell} S_\ell$ shown for $\epsilon_\phi$ values close to the resonant points $(\epsilon_\phi^\star)_{n\ell}$ for $n=3$  and $\ell=0,1,2$ (upper-right, bottom-left, bottom-right panels, resp.). Lighter curves correspond to $\epsilon_\phi$ values closer to $(\epsilon_\phi^\star)_{n\ell}$. 
    While the $s$-wave resonances are nearly symmetric under small variations around the exact resonant point, here demonstrated for $(\epsilon_\phi^\star)_{30} = 0.069725$  (upper-right), the higher partial waves (lower panels) show a velocity spike for $\epsilon_\phi \searrow (\epsilon_\phi^\star)_{n \ell}$ (solid) but not for $\epsilon_\phi \nearrow (\epsilon_\phi^\star)_{n \ell}$ (dashed). 
    The spikes are a characteristic feature of the presence of a quasi-bound state, whose discrete positive energy (inverse lifetime) sets the peak location (width of the peak). 
    While the resonant $s$-wave Sommerfeld factor scales as $S_0 \propto \epsilon_v^{-2}$ for $\epsilon_v \lesssim \epsilon_\phi$, all higher partial waves  close to resonance exhibit the $S_{\ell \geq 1} \propto \epsilon_v^{-4}$ anomalous scaling.}
    \label{fig:epsphiv}
\end{figure}

We consider a system of two particles of mass $m_\chi$ which interact via an attractive Yukawa potential of range $1/m_\phi$. In dimensionless variables $x \equiv \alpha m_\chi r$, the time-independent, radially reduced Schr\"odinger equation for this system is given by
\begin{align}
 \label{eq:SchroEq1}
    -u''_{\ell}(x)+V_{\ell}^{\text{eff}}(x)u_{\ell}(x)=\epsilon^2 u_{\ell}(x) \;,   
\end{align}
with the effective potential
\begin{equation}
    \label{eq:EffPot}
    V_{\ell}^{\text{eff}} \equiv -\frac{e^{-\epsilon_\phi x}}{x}+\frac{\ell(\ell+1)}{x^2}\,.
\end{equation}
The dimensionless energy variable is defined as $\epsilon^2 \equiv E/(\alpha^2 m_\chi)$, and $\epsilon$ equals $\epsilon_v$ defined in Eq.~(\ref{eq:epsv_def}) for positive kinetic energy $E=m_\chi^2 v^2/4$ of the relative motion. The dimensionless energy of a bound state will be denoted  by $\epsilon_{n\ell}^2$, which approaches the degenerate Coulomb value $\epsilon_{n\ell}^2 \rightarrow -1/(4n^2)$ in the limit $\epsilon_\phi \rightarrow 0$.
We follow standard practice and solve Eq.~(\ref{eq:SchroEq1}) numerically for positive energies. In particular, the same boundary conditions as in scattering theory are adopted, i.e. $u_\ell(x)$ is regular at the origin and $u_{\ell}(x) \rightarrow \sin(\epsilon_v x -\frac{\ell\pi}{2}+\delta_{\ell})$ for $x \rightarrow \infty$. With these, the Sommerfeld factor is related to the reduced wave-function as (e.g., Ref.~\cite{Slatyer:2009vg}):
\begin{equation}
    \label{eq:SEl}
    S_{\ell} =\left|\frac{(2\ell+1)!! \, \partial_x^{\ell+1}u_\ell(0)}{(\ell+1)! \, \epsilon_v^{\ell+1}}\right|^{\,2}.
\end{equation}
In order to understand the resonant structure of the Sommerfeld factor 
as a function of the mediator mass, resp. $\epsilon_\phi$, we first examine the negative-energy solutions of Eq.~(\ref{eq:SchroEq1}). Unlike in the Coulomb case, the Schr\"odinger equation with effective Yukawa potential in~\eqref{eq:EffPot} has only a finite number of bound-state solutions. 
When $\epsilon_\phi$ is too large, the potential is too short-ranged to 
sustain a bound state. For $\ell=0$ and 
$(\epsilon_\phi^\star)_{10}\equiv 0.5953\ldots$ 
the first ($n=1$) bound-state appears, which aligns with the position of the first resonant enhancement of the annihilation cross 
section. As $\epsilon_\phi$ is decreased, the $n>1$ and $\ell\geq 0$  bound states appear and at the critical values $(\epsilon_\phi^\star)_{n \ell}$ of $\epsilon_\phi$, where the corresponding bound state has zero energy, 
DM annihilation is resonantly enhanced. The sequence of resonance 
peaks is illustrated in the upper-left panel of Fig.~\ref{fig:epsphiv} below 
for the partial waves $\ell=0,1,2$. 


\begin{figure}[t]
    \centering
    \includegraphics[scale=0.85]{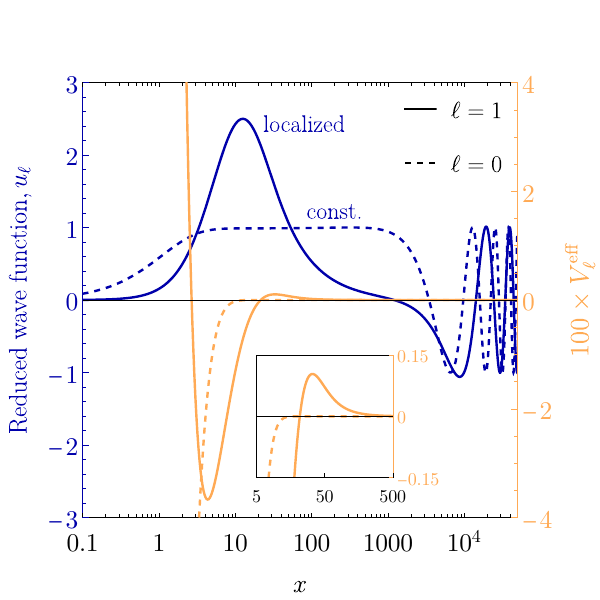}
    \caption{\small Reduced wave function (blue) and effective potential (orange), demonstrating qualitative differences between the first $p$-wave (solid) and $s$-wave (dashed) resonance. In both cases, the resonance is approached from above, i.e. $\epsilon_\phi \gtrsim (\epsilon_{\phi}^\star)_{n\ell}$, where for $\ell=0$, we show $\epsilon_\phi = 0.595307 \gtrsim (\epsilon_\phi^\star)_{10}$ and, for $\ell=1$, $\epsilon_\phi= 0.110109 \gtrsim (\epsilon_\phi^\star)_{21}$. The inset highlights the presence of the potential barrier for $\ell=1$.
    The velocity parameter has been chosen to be $\epsilon_v = 3 \times 10^{-4} \approx \epsilon_{21}$, the QBS energy. }
    \label{fig:qb}
\end{figure}


A key observation is that in the vicinity of the critical values 
there is a marked difference between the well-known $\ell=0$ 
resonances and the higher partial-waves when $\epsilon_\phi$ is slightly above $(\epsilon_\phi^\star)_{n \ell}$. This is related to the fact 
that for $\ell \geq 1$ the effective potential presents a centrifugal barrier as is shown in Fig.~\ref{fig:qb} (and highlighted inset). Hence, while for $\ell=0$ the bound-state simply disappears from the spectrum as $\epsilon_\phi$ crosses $(\epsilon_\phi^\star)_{n \ell}$, for $\ell \geq 1$ the presence of this barrier 
leads to a \emph{metastable} state with small positive energy $0<\epsilon_{n\ell}^2 \ll 1$ and finite lifetime $\gamma_{n\ell}$.\footnote{This terminology implies a semi-classical interpretation. The ``metastable state" correspond to a pole in the complex energy plane.} Following \cite[§134]{Landau:1991wop}, we refer to it as a \emph{quasi-bound state}.

We now analyze the wave functions $u_\ell(x)$ for $\epsilon_\phi \gtrsim (\epsilon_\phi^\star)_{n \ell}$ in the $\ell=0$ and $\ell \geq 1$ cases. In Fig.~\ref{fig:qb} we display $u_{0}(x)$  of a $\ell=0$ scattering state with small energy $0<\epsilon_v \ll 1$ for $\epsilon_\phi \gtrsim (\epsilon_\phi^\star)_{10}$ (dashed blue). We note that the wave function is non-localized at $x \lesssim \frac{1}{\epsilon_v}$  due to its scaling $u_{0}(x) \sim e^{-\epsilon_{n0} x}$ with $0<\epsilon_{n0} \ll 1$ in this region  
 \cite[§133]{Landau:1991wop}. Outside this region, for $x \gtrsim \frac{1}{\epsilon_v}$, the oscillatory behavior expected for a scattering state starts. We compare this to the  wave function of the $\ell=1$ scattering state, $u_{1}(x)$, at energy close to the quasi-bound state energy $\epsilon_v \approx \epsilon_{21} \ll 1$ (solid blue). Different from the $s$-wave case, the presence of the centrifugal barrier suppresses the wave function. In fact, in this region defined by \begin{equation}
 -\frac{W_{-1}\left(-\ell(\ell+1)\epsilon_\phi\right)}{\epsilon_\phi} \lesssim x \lesssim \frac{1}{\epsilon_v}\,,
 \end{equation}
 where the lower limit is the second zero of the effective potential, where the potential becomes positive,\footnote{Here $W_{-1}(z)$ denotes the second branch of the Lambert $W$ function.}
the centrifugal term is dominant compared to the energy and the Yukawa potential, and the wave function scales as  $u_{\ell}(x) \sim x^{-\ell}$. For $x \gtrsim \frac{1}{\epsilon_v}$, the oscillatory behavior starts. 

The existence of a QBS for $\ell\geq 1$ has important implications for the velocity dependence of the Sommerfeld factor when $\epsilon_\phi$ is close to 
a critical value. First, let us recall that for the familiar $s$-wave case, 
when $\epsilon_\phi$ is not close to a critical value $S_0 \propto 1/v$ in the velocity region $\epsilon_v > \epsilon_\phi$ and  saturates to a constant in the small-velocity limit $\epsilon_v < \epsilon_\phi$. On the other hand, 
close to a resonance value $(\epsilon_\phi^*)_{n0}$, $S_0$ exhibits the power law behavior $S_0 \propto \epsilon_v^{-2}$ in the velocity region $\epsilon_v > |\epsilon_{n0}|$ and it saturates to a constant value for $\epsilon_v < |\epsilon_{n0}|$. This behavior is observed regardless of whether $\epsilon_\phi \gtrsim (\epsilon_\phi^\star)_{n0}$ or $\epsilon_\phi \lesssim (\epsilon_\phi^\star)_{n0}$, i.e. both for negative $\epsilon_{n0}^2$ (small energy bound-state in the spectrum) and for positive $\epsilon_{n0}^2$ (no bound-state), and shown  in the upper-right panel of Fig.~\ref{fig:epsphiv}. In contrast to these well-known features, the $\ell \geq 1$ Sommerfeld factors exhibit an anomalous velocity scaling and a resonance in the {\em velocity} dependence near the critical 
quasi-bound values $(\epsilon_\phi^*)_{n\ell}$. These features are demonstrated in the lower panels of Fig.~\ref{fig:epsphiv} for $\ell=1,2$ and will be discussed in the following.


For $\epsilon_\phi \rightarrow (\epsilon_\phi^\star)_{n \ell}$ from above or below, and in the range $|\epsilon_{n\ell}| \ll \epsilon_v \ll \epsilon_\phi$, the Sommerfeld factor exhibits the anomalous scaling $S_\ell \propto \epsilon_v^{-4}$ for all $\ell \geq 1$, which differs from the resonant $s$-wave scaling $S_0 \propto \epsilon_v^{-2}$, while for even smaller velocities $\epsilon_v \ll |\epsilon_{n\ell}|$, it saturates to a constant value, analogously to the $s$-wave case. Combined with the $\epsilon_v^{2\ell}$ factor from the short-distance partial-wave cross section, this leads to the velocity dependence displayed in the upper-right and lower panels. 

The anomalous scaling has also been noted in Ref.~\cite{Kamada:2023iol}, based on the Watson theorem, although the connection to quasi-bound states 
was not made. It was pointed out that, due to this scaling, partial-wave unitarity can become an issue also for $p$-wave annihilation, while higher partial waves  do not violate unitarity due to the $\epsilon_v^{2 l}$ suppression in Eq.~(\ref{eq:ldecomposition}). It is worth to add that for $\epsilon_\phi \searrow  (\epsilon_\phi^\star)_{n \ell}$ partial-wave unitarity violation can be problematic for \emph{any} partial wave due to the additional resonant enhancement from the quasi-bound state contribution discussed next.

When the parametric conditions for the QBS are met, i.e. for $\ell\geq1$ and $\epsilon_\phi \searrow (\epsilon_\phi^\star)_{n \ell}$, there occur spikes in the velocity dependence of the Sommerfeld factor. The peak position equals the discrete positive value of the QBS energy $\epsilon_{nl}$, while the width of the spike is determined by the inverse QBS lifetime. As clearly visible, this needs to be distinguished from $\epsilon_\phi \nearrow (\epsilon_\phi^\star)_{n \ell}$, where no such spike exists at positive energies but instead a weakly coupled bound state. The asymmetric behavior when approaching $(\epsilon_\phi^\star)_{n \ell}$ from above or below is absent in the $s$-wave case.

The spike in the velocity spectrum is similar to the spike that appears when two dark matter particles annihilate through an 
$s$-channel resonance with a mass slightly above $2 m_\chi$. 
The $\ell >1$ QBS situation therefore realizes the ``super-resonant" enhancement discussed in \cite{Beneke:2022rjv} from the coincidence of Sommerfeld and resonant enhancement---with the intriguing difference that the resonance is not another particle but a metastable dark-matter bound state generated dynamically by the Yukawa potential itself.


\subsection{Quasi-bound states in the wino model}
\label{sec:Quasi-bound_states_in_the_wino_model}

We briefly discuss the potential appearance of quasi-bound states in a multi-channel Schr\"odinger equation and consider the minimal spin-$\frac{1}{2}$ electroweak dark-matter triplet (``wino") model. The relevant particles are the lightest neutralino $\chi^0$ with mass $m_\chi$  and its charged partners $\chi^\pm$ with mass $m_{\chi}+\delta m$. We are interested in the Sommerfeld factor for $\chi^0 \chi^0$ annihilation into SM particles, which differs from unity only for the even $\ell+s$ states \cite{Beneke:2014gja}. In dimensionless variables, the potential is \cite{Hisano:2004ds} 
\begin{equation}
    \label{eq:winoPot}
    V^{W,\text{eff}}_{\ell}(x)=
    \begingroup
    \setlength\arraycolsep{10pt}
    \begin{pmatrix}
        \frac{\ell(\ell+1)}{x^2} & - \frac{\sqrt{2}\,e^{-c_w \epsilon_z x}}{x} \\
        - \frac{\sqrt{2}\,e^{-c_w \epsilon_z x}}{x} & -\frac{s_w^2}{x} -  c_w^2 \frac{e^{-\epsilon_z x}}{x} +\frac{\ell(\ell+1)}{x^2}+2\epsilon_{\delta m}
    \end{pmatrix}\,,
    \endgroup
\end{equation}
where $\epsilon_z \equiv m_Z/(\alpha_2 m_{\chi})$, $\epsilon_{\delta m} \equiv \delta m/(\alpha_2^2 m_{\chi})$.  $s_w$, $c_w$ denote the sine / cosine of the Weinberg angle. The effective potential \eqref{eq:winoPot} includes a long-range Coulomb term in the diagonal chargino channel, which potentially erases the centrifugal barrier. The presence or absence of the centrifugal barrier is determined by two factors: the relative strength between the short-range and the long-range terms, which is related to the Weinberg angle $\theta_w$, and whether the total energy is sufficient to produce a real chargino pair, i.e. by the relative magnitude of $\epsilon_v^2$ and $\epsilon_{\delta m}$.

First, we focus on the chargino potential
\begin{equation}
    \label{eq:winoPot_Y+C}
    V^{Y+C,\text{eff}}_{\ell} \equiv -\frac{s_w^2}{x}-c_w^2 \frac{e^{-\epsilon_z x}}{x}+\frac{\ell(\ell+1)}{x^2}.
\end{equation}
At fixed $\epsilon_z$, by increasing the Weinberg angle, the centrifugal barrier is erased and the effective potential becomes purely attractive. The condition for the presence of a centrifugal barrier is $\cot(\theta_w)>e$, which is not satisfied by the Standard Model value.
Second, another short-range contribution to the potential comes from the off-diagonal neutralino-chargino interaction. Taking into account also this contribution is still not sufficient to create a centrifugal barrier. Third, the mass splitting term plays a key role. In fact, for $\epsilon_v^2<\epsilon_{\delta m}$, the production of a chargino pair is kinematically forbidden and the chargino channel is closed. Therefore, the effect of the diagonal chargino term \eqref{eq:winoPot_Y+C} is strongly suppressed and the main contribution comes from the off-diagonal Yukawa terms, which are short-range, hence a centrifugal barrier is formed. Adopting $\alpha=1/128.943$, $m_W=80.385$~GeV and $m_Z=91.1876$~GeV and the mass splitting $\delta m=164.1$~MeV, the first $p$-wave resonance occurs at $m_{\chi} \approx 11~\text{TeV}$, in which case the chargino channel is closed for $v<0.011$. 
Therefore, quasi-bound state resonances centered at $\epsilon_{n\ell}^2<\epsilon_{\delta m}$ will not be erased by the Coulomb term. This allows for the presence of 
a ${}^3P_J$ quasi-bound state in the wino model, as will be demonstrated in Sec.~\ref{sec:wino_model}.


\section{WKB approximation of the quasi-bound state Sommerfeld factor}
\label{sec:WKB}

Returning to the single-channel Yukawa potential, we exploit the WKB method in order to estimate the Sommerfeld factor for $\ell \geq 1$ in the presence of a quasi-bound state and to obtain an analytical understanding of the key features. The motion of a scattering particle in the effective potential $V_{\ell}^{\text{eff}}(x)$ with energy lower than the maximum of the centrifugal barrier has three classical turning points, which will be denoted as $x_1<x_2<x_3$ and which identify two classically accessible and two inaccessible regions.
The condition for the application of the WKB method reads 
\begin{equation}
    \label{eq:WKBcondition}
    \left|\frac{dp(x)}{dx}\right| \ll p^2(x),
\end{equation}
where $p(x) \equiv \sqrt{V_{\ell}^{\text{eff}}(x)-\epsilon_v^2}$ for $x \in [0,x_1]\cup[x_2,x_3]$ and $p(x) \equiv \sqrt{\epsilon_v^2-V_{\ell}^{\text{eff}}(x)}$ for $x \in [x_1,x_2]\cup[x_3,\infty]$. This condition in the classically inaccessible region leads to $\sqrt{\ell(\ell+1)} \gg 1$, in accordance with the correspondence principle. We employ the Langer modification \cite{Langer:1937qr}, which is expected to improve the WKB approximation particularly at lower $\ell$ values. It consists in the change of variable $y \equiv \ln x$ and in the use of the wave function $U_{\ell}(y) \equiv e^{-y/2}u_{\ell}(e^y)$, which fulfills
\begin{equation}
    -\frac{d^2U_{\ell}(y)}{dy^2}+P^2(y) U_{\ell}(y)=0,
\end{equation}
where $P^2(y) \equiv e^{2y} \left([p(e^y)]^2-1/4\right)$.
This equation for $U_{\ell}(y)$ is then solved by means of the WKB approximation.
By transforming back the approximate solution for $U_\ell(x)$ to the original function, the so obtained wave function $u_{\ell}(x)$ is identical to the standard WKB approximation, but for the potential with $\ell(\ell+1) \mapsto (\ell+1/2)^2$:
\begin{equation}
V_{\ell}^{\text{Lan}} \equiv -\frac{e^{-\epsilon_\phi x}}{x}+\frac{(\ell+1/2)^2}{x^2},
\end{equation}
which should hence be used in the definition of the classical momentum $p(x)$. 

In order to compute the Sommerfeld factor \eqref{eq:SEl} in the WKB approximation for a quasi-bound state resonance, one first follows the standard WKB method and 
glues together the approximations for a scattering state with 
asymptotic behavior $u_{l}(x) \to \sin(\epsilon_v x -\frac{\ell \pi}{2}+\delta_\ell)$ and energy below the height of the barrier in the two classically accessible  and two classically inaccessible regions ($x<x_1$ and $x_2 <x<x_3$), see Fig.~\ref{fig:qb}.   Once the Sommerfeld factor has been obtained, it is necessary to add the information on the presence of a quasi-bound state in the spectrum. Differently from a scattering state, a metastable state requires a progressive wave $u_{l} \propto e^{i \epsilon_v x}$ as a boundary condition for $x \to \infty$ with complex $\epsilon_v$. This leads to a complex quantization condition, which allows us to estimate the energy $\epsilon_{n\ell}$ (real part) and the width $\gamma_{n\ell}$ (imaginary part) of the quasi-bound state. The last requirement is that the velocity of the considered scattering state $\epsilon_v$ is close to the quasi-bound state energy $\epsilon_{n\ell}$, leading to the final result:
\begin{eqnarray}
S_{\ell} &\approx& \frac{[(2\ell+1)!!]^2}{(2\ell+1) T_{n\ell}} \,\epsilon_{n\ell}^{-2\ell-1}\frac{(\gamma_{n\ell}/2)}{(\epsilon_v^2-\epsilon_{n\ell}^2)^2+(\gamma_{n\ell}/2)^2}
\nonumber\\
 && \times \exp \left\{2 \int_0^{x_1} dx' \, \ln(x') \frac{d}{dx'}[x'p(x')]\right\}\,,
\label{S_WKB}
\end{eqnarray}
where the exponent is evaluated at $\epsilon_v^2=\epsilon_{n \ell}^2$. 
The details of the derivation are presented in App.~\ref{app:WKB_details}. 
In the above expression, $T_{n\ell}$ is defined as the classical oscillation period in the well (see \eqref{period}). Eq.~\eqref{S_WKB} provides a non-trivial extension of the WKB treatment of bound-state annihilation~\cite{POPOV:1995} to the decay of metastable states. 

\begin{figure}[t]
    \centering
    \includegraphics[scale=0.88]{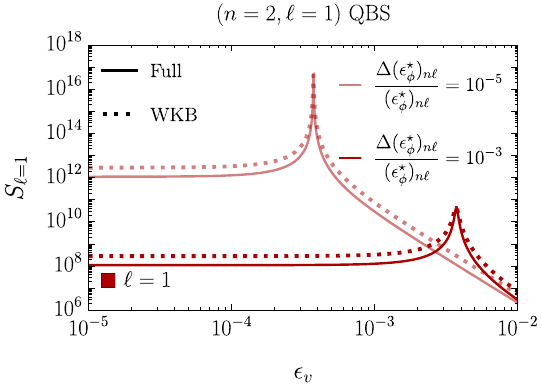}
    \includegraphics[scale=0.88]{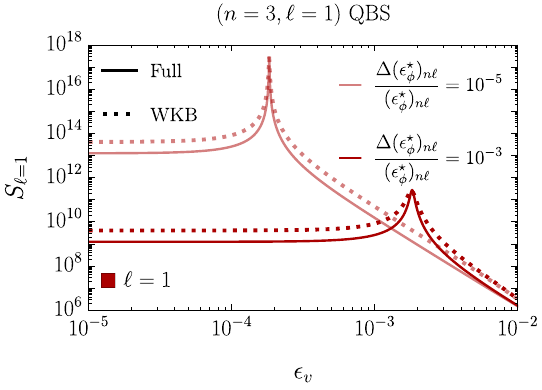} \\
    \vspace{0.5 cm}
    \includegraphics[scale=0.88]{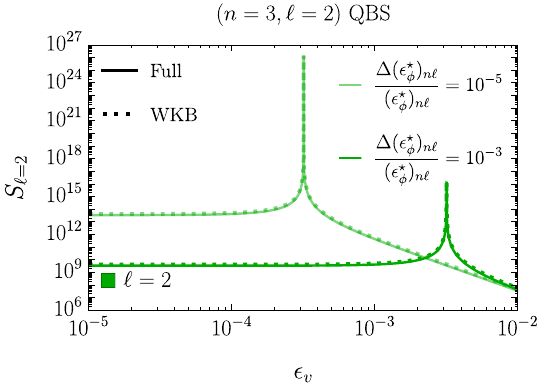}
    \includegraphics[scale=0.88]{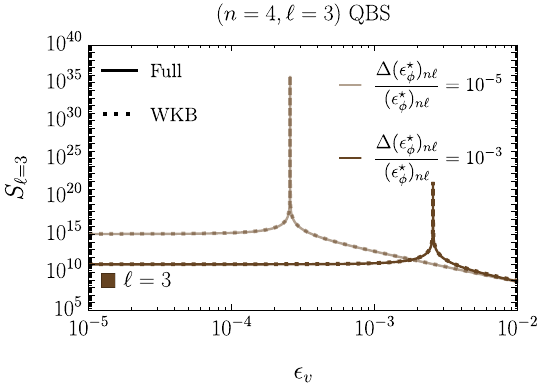}
    \caption{\small Comparison between the Sommerfeld factors obtained from i) a numerical approach to solve the  Schr\"odinger equation with an attractive Yukawa potential (``Full"), and ii) the corresponding WKB result~(\ref{S_WKB}). The $\epsilon_\phi$ parameter entering the WKB estimate is chosen such that the (dimensionless) QBS energy coincides with the peak location of the full result. While $S_{\ell=1}$ in the upper panel differs at most by a factor 3 for $n=2$ (left) and $n=3$ (right), the quality of the WKB approximation improves as expected towards higher partial waves (lower panels) where the Sommerfeld factors differ in the saturated region by about $22$\% for $n=3,\ell=2$ (left), and 4\% for $n=4,\ell=3$ (right). Lighter curves correspond to $\epsilon_\phi$ values closer to $(\epsilon_\phi^\star)_{n\ell}$, see \eqref{eq:Adef}.}
    \label{fig:Full_VS_WKB}
\end{figure}

In expression \eqref{S_WKB} one can recognize the Breit-Wigner line-shape, which confirms that the quasi-bound state acts akin to a resonance in the velocity-dependence of the Sommerfeld factor. We highlight the fact that the classical oscillation period, width and the exponential factor depend only on the QBS energy $\epsilon_{nl}$. Therefore, the velocity dependence is set by the term $(\epsilon_v^2-\epsilon_{n\ell}^2)^2$, which implies $S_{\ell} \propto \epsilon_v^{-4}$ for $\epsilon_v \gg \epsilon_{n\ell}$ and $S_{\ell} \to \text{const}$ for $\epsilon_v \ll \epsilon_{n\ell}$ for any $\ell \geq 1$. These observations already explain all our earlier findings in Sec.~\ref{sec:Quasi-bound_states} qualitatively. In fact, the anomalous scaling, the Breit-Wigner peak at $\epsilon_v \sim \epsilon_{n\ell}$, and saturation obtained in the WKB approximation are in agreement with the fully numerical solution of the Schr\"odinger equation \eqref{eq:SchroEq1} (referred to as ``full'' in the following), as demonstrated in Fig.~\ref{fig:Full_VS_WKB}.


\begin{table}[t]
\centering
\begin{tabular}{c"c|ccc}
& \multicolumn{1}{c|}{$n=4$} & \multicolumn{1}{c|}{$n=3$} & \multicolumn{1}{c|}{$n=2$} & \multicolumn{1}{c|}{$n=1$} \\ \thickhline
$\ell=0$ & \multicolumn{1}{c|}{$\bf 0.03941$} & \multicolumn{1}{c|}{$\bf 0.06973$} & \multicolumn{1}{c|}{$\bf 0.1551$} & \multicolumn{1}{c|}{$\bf 0.5953$} \\ \hline
$\ell=1$ & \multicolumn{1}{c|}{\begin{tabular}[c]{@{}l@{}}$\bf 0.03394$  \\ $ 0.03340$\end{tabular}} & \multicolumn{1}{c|}{\begin{tabular}[c]{@{}l@{}}$\bf 0.05636$  \\ $0.05515$\end{tabular}} & \multicolumn{1}{c|}{\begin{tabular}[c]{@{}l@{}}$\bf 0.1101$\\ $0.1067$\end{tabular}} & \\ \cline{1-4}
$\ell=2$ & \multicolumn{1}{c|}{\begin{tabular}[c]{@{}l@{}}$\bf 0.02905$\\$0.02877$\end{tabular}} & \multicolumn{1}{c|}{\begin{tabular}[c]{@{}l@{}}$\bf 0.04567 $\\$0.04512$\end{tabular}} & & \\ \cline{1-3}
$\ell=3$ & \multicolumn{1}{c|}{\begin{tabular}[c]{@{}l@{}}$\bf 0.02492$\\$0.02476$\end{tabular}} & & & \\ \cline{1-2}
\end{tabular}
\caption{\small Critical values $(\epsilon_\phi^\star)_{n\ell}$ obtained numerically (bold) and from our WKB approximation. The largest difference ($3.1\%$) appears for $n=2$ and $\ell=1$.}
\label{tab:critical}
\end{table}


We turn now to a quantitative comparison between the WKB approximation and the full numerical solution for the corresponding quantities. Our starting point considers the critical values $(\epsilon_\phi^\star)_{n\ell}$ which lead to the anomalous velocity scaling of the Sommerfeld factor for $\ell \geq 1$. These values are reported in Tab.~\ref{tab:critical} for both the numerical results (bold) and the WKB approximation ($\ell>0$ only) within four-digit precision. The critical values in the latter method are obtained from the solution of the quantization condition \eqref{real_quantization} for vanishing QBS energies $\epsilon_{n\ell}=0$, consistent with the definition of $(\epsilon_\phi^\star)_{n\ell}$. The expected higher quality of the WKB approximation for increasing $\ell$ can be recognized.

The WKB results make simple predictions in the regime $\epsilon_{n\ell} \ll 1$ which we now compare against the full result. In particular, the solution to the quantization condition \eqref{real_quantization} relates the QBS energy and the $\epsilon_\phi$ parameter. By numerical evidence, the solution to the quantization condition in the regime $\epsilon_{n\ell} \ll 1$ suggests the parametrization
\begin{equation}
    \epsilon_{n\ell}^2=A_{n\ell} \frac{\Delta (\epsilon_\phi^\star)}{(\epsilon_\phi^\star)_{n\ell}},
    \label{eq:Adef}
\end{equation}
where we introduced $\Delta (\epsilon_\phi^\star)_{n\ell} \equiv \epsilon_\phi - (\epsilon_\phi^\star)_{n\ell}$ and the constant parameter $A_{n\ell}$. We test the quality of this prediction against the full result. To do so, it is assumed that the QBS peak location in the full result can be equated with the QBS energy on the left-hand side of \eqref{eq:Adef}. The value of $A_{n\ell}$ is then fixed at the choice of $\Delta (\epsilon_\phi^\star)_{n\ell}/(\epsilon_\phi^\star)_{n\ell}=10^{-3}$ and reported in Tab.~\ref{tab:Full_VS_WKB}.  Using the same value of $A_{n\ell}$ also for $\epsilon_\phi$ values closer to the critical one, the predicted energy agrees with the peak location of the full result at the $1 \%$ level (tested for $\Delta (\epsilon_\phi^\star)_{n\ell}/(\epsilon_\phi^\star)_{n\ell}=10^{-5}$ and all $n, \ell$ values considered in Tab.~\ref{tab:Full_VS_WKB}). The scaling \eqref{eq:Adef} can also be noticed in Fig.~\ref{fig:Full_VS_WKB}.


\begin{table}[b]
\begin{center}
\footnotesize
\begin{tabular}{|c|cccccccc|}
\hline
\multirow{2}{*}{$(n,\ell)$} & \multicolumn{8}{c|}{$\Delta (\epsilon_\phi^\star)_{n\ell}/(\epsilon_\phi^\star)_{n\ell}=10^{-3}$} \\ \cline{2-9}
& \multicolumn{1}{c|}{$A_{n\ell}$} & \multicolumn{1}{c|}{$A_{n\ell}^{\text{WKB}}$} & \multicolumn{1}{c|}{$B_{n\ell}$} & \multicolumn{1}{c|}{$B_{n\ell}^{\text{WKB}}$} & \multicolumn{1}{c|}{$C_{n\ell}$} & \multicolumn{1}{c|}{$C_{n\ell}^{\text{WKB}}$} & \multicolumn{1}{c|}{$\epsilon_\phi$} & $\epsilon_\phi^\text{WKB}$ \\ \hline
$(2,1)$ & \multicolumn{1}{c|}{$0.0139$} & \multicolumn{1}{c|}{$0.0222$} & \multicolumn{1}{c|}{$0.0209$} & \multicolumn{1}{c|}{$0.0536$} & \multicolumn{1}{c|}{$26.4$} & \multicolumn{1}{c|}{$44.3$} & \multicolumn{1}{c|}{$0.110219$} & $0.106793$ \\ \hline
$(3,1)$ & \multicolumn{1}{c|}{$0.00333$} & \multicolumn{1}{c|}{$0.00563$} & \multicolumn{1}{c|}{$0.0138$} & \multicolumn{1}{c|}{$0.0441$} & \multicolumn{1}{c|}{$74.6$} & \multicolumn{1}{c|}{$141$} & \multicolumn{1}{c|}{$0.0564116$} & $0.0551847$ \\ \hline
$(3,2)$ & \multicolumn{1}{c|}{$0.0101$} & \multicolumn{1}{c|}{$0.0115$} & \multicolumn{1}{c|}{$0.340$} & \multicolumn{1}{c|}{$0.417$} & \multicolumn{1}{c|}{$2.93 \times 10^4$} & \multicolumn{1}{c|}{$3.33 \times 10^4$} & \multicolumn{1}{c|}{$0.0457182$} & $0.0451576$ \\ \hline
$(4,3)$ & \multicolumn{1}{c|}{$0.00653$} & \multicolumn{1}{c|}{$0.00692$} & \multicolumn{1}{c|}{$4.80$} & \multicolumn{1}{c|}{$5.03$} & \multicolumn{1}{c|}{$6.61 \times 10^7$} & \multicolumn{1}{c|}{$6.90 \times 10^7$} & \multicolumn{1}{c|}{$0.0249405$} & $0.0247786$ \\ \hline
\end{tabular}
\caption{\small Coefficients $A_{n\ell}$, $B_{n\ell}$ and $C_{n\ell}$ of the parametrization \eqref{S_parametrization} for the full results and for the WKB approximation. We also report the precise values of the $\epsilon_\phi$ which have been chosen.}
\label{tab:Full_VS_WKB}
\end{center}
\end{table}


In App.~\ref{app:WKB_details} we show that in the same regime $\epsilon_{nl} \ll 1$, the width can be parameterized as $\gamma_{n\ell}  = C_{n\ell} \epsilon_{n\ell}^{2\ell+1}$, while $T_{n \ell}$ and the exponent in \eqref{S_WKB} approach finite values. Consequently, the WKB result \eqref{S_WKB} suggests the following parametrization of the Sommerfeld factor:
\begin{equation}
    \label{S_parametrization}
    S_{\ell}=\frac{B_{n\ell}}{(\epsilon_v^2-\epsilon_{n\ell}^2)^2+C_{n\ell}^2 \epsilon_{n\ell}^{4\ell+2}/4}\,,
\end{equation}
with constant coefficients $B_{n\ell}$ and $C_{n\ell}$ and energy given by \eqref{eq:Adef}. By matching the full results to this parametrization  consistently for $\Delta (\epsilon_\phi^\star)_{n\ell}/(\epsilon_\phi^\star)_{n\ell}=10^{-3}$, we obtain the values of $B_{n\ell}$ and $C_{n\ell}$ reported in Tab.~\ref{tab:Full_VS_WKB}. Concretely, we determine $B_{n\ell}$ in the limit $\epsilon_v \rightarrow 0$, where $S_\ell \approx B_{n \ell}/\epsilon_{n \ell}^4$, and use this value to determine $C_{n\ell}$ at $\epsilon_v=\epsilon_{n \ell}$.
In this way, the matched parametrization reproduces the full numerical results at $\mathcal{O}(1\%)$ in the range $\epsilon_v \in \left[10^{-5},10^{-2}\right]$, which implies that the WKB approximation correctly predicts the Breit-Wigner shape and the velocity dependence of the Sommerfeld factor.
The parameters $B_{n\ell}$ and $C_{n\ell}$ also exhibit only a minor  dependence on $\Delta (\epsilon_\phi^\star)_{n\ell}/(\epsilon_\phi^\star)_{n\ell}$ ($ \leq 2\%$ deviation for $\Delta (\epsilon_\phi^\star)_{n\ell}/(\epsilon_\phi^\star)_{n\ell}=10^{-5}$ and all $n, \ell$ considered in Tab.~\ref{tab:Full_VS_WKB}).

Similarly, the coefficients in \eqref{eq:Adef} and \eqref{S_parametrization} are matched to their values in the WKB approximation. The so obtained $A_{n\ell}^{\text{WKB}}$, $B_{n\ell}^{\text{WKB}}$, and $C_{n\ell}^{\text{WKB}}$ are also listed in Tab.~\ref{tab:Full_VS_WKB}. They share a similar minor dependence on the matching choice of the QBS energy. In more detail, we first adjust the $\epsilon_{n \ell}$ of the WKB approximation to coincide with the peak location of the full numerical results for $\Delta (\epsilon_\phi^\star)_{n\ell}/(\epsilon_\phi^\star)_{n\ell}=10^{-3}$ to make the spike locations coincide. 
From the solution of the quantization condition for this $\epsilon_{n \ell}$, we obtain $\epsilon_\phi^{\text{WKB}}$, which determines the reported $A_{n\ell}^{\text{WKB}}$ value. The expression \eqref{width} for the width is used to determine $C_{n\ell}^{\text{WKB}}$ directly, and $B_{n\ell}^{\text{WKB}}$ from \eqref{S_parametrization} in the low velocity limit. As expected, the coefficients obtained from the full result and WKB approximation agree better for higher $\ell$.


\section{Examples}
\label{sec:Examples}

In this section we illustrate the phenomenological impact of QBS on indirect detection signals, choosing representative dark matter models featuring a light mediator, one in which annihilation is dominated by the $p$-wave, one in which both $s$- and $p$-wave contribute, as well as wino dark matter.


\subsection{Scalar mediator model}
\label{sec:Scalar_mediator_model}

As our first example, we consider a dark sector composed of a fermionic dark matter particle $\chi$, with mass $m_\chi$, and a light \emph{scalar} with mass $m_\phi$ coupled to dark matter via the  Yukawa interaction (see e.g.~\cite{Kaplinghat:2013yxa,Kainulainen:2015sva,Kahlhoefer:2017umn,Hufnagel:2018bjp,Hambye:2019tjt, Biondini:2021ccr, Biondini:2023ksj, Chen:2024njd}),
\begin{equation}
 \mathcal{L} = \Bar{\chi}(i \slashed{\partial}-m_\chi)\chi+\frac{1}{2}\partial_\mu \phi \partial^\mu \phi -\frac{1}{2}m_\phi^2 \phi^2 - g_\phi \bar{\chi} \chi \phi\,.
\end{equation}
The dark matter particles $\chi$ can annihilate into two mediators $\Bar{\chi} \chi \to \phi \phi$.  The leading contribution to the annihilation cross section is $p$-wave and reads
\begin{equation}
    \sigma v = \frac{3\pi \alpha^2}{8 m_\chi^2} v^2 S_{\ell=1}\,,
\end{equation}
with $\alpha \equiv g_\phi^2/(4\pi)$. As an illustrative example for the impact of the QBS, we choose $\epsilon_\phi=0.1102 \gtrsim (\epsilon_\phi^\star)_{21}$ close to the first $p$-wave resonance, as well as a DM mass of $m_\chi=75\, \text{GeV}$ and $\alpha=0.01103$. Using~\eqref{eq:epsilonphi} the mediator mass is also fixed,  $m_\phi=91.16 \,\text{MeV}$. This choice leads to a relic abundance via freeze-out\footnote{We use the DRAKE code~\cite{Binder:2021bmg} to compute the relic abundance.} of $\Bar\chi\chi\to\phi\phi$ in agreement with the Planck measurement~\cite{Planck:2018vyg}, assuming that the dark sector evolves with the same temperature as the SM thermal bath. The latter property can be ensured by a weak coupling of the mediator to the SM that also leads to a decay of $\phi$ into SM particles. Since the details of this coupling are not essential here, we assume for simplicity a decay $\phi\to e^+e^-$ with lifetime such that $\phi$ is effectively stable during the annihilation process but decays well before Big Bang Nucleosynthesis (BBN).\footnote{The coupling to the SM could for example arise from a mixing of $\phi$ with the Higgs~\cite{Kaplinghat:2013yxa}. However, the resulting coupling to quarks leads to stringent constraints from direct detection, enforcing a mediator lifetime in conflict with BBN bounds, except for sufficiently heavy masses~\cite{Hufnagel:2018bjp}. We note that these constraints are greatly relaxed when assuming an effective coupling of $\phi$ to $e^+e^-$ only, which is sufficient for the phenomenology discussed here. For other possibilities we refer to~\cite{Hambye:2019tjt}.}

The cascade annihilation process $\Bar{\chi} \chi \to \phi \phi$ with subsequent mediator decay $\phi\to e^+e^-$ gives rise to indirect detection signatures from dwarf spheroidal galaxies (dSph) and from the galactic center (GC), which are constrained by positron flux measurements from AMS-02~\cite{AMS:2013fma,AMS:2021nhj} and gamma-ray observations from e.g. Fermi-LAT, MAGIC, H.E.S.S. and the future CTA~\cite{Fermi-LAT:2015att, MAGIC:2022acl, HESS:2016mib, Eckner:2021jow}.
To derive indirect detection constraints, the cross section needs to be averaged over the velocity distribution within the respective targets, which is of particular relevance in view of the spike produced by the QBS. For illustration, we provide estimates following the standard practice of assuming a Maxwellian distribution, using the benchmark values $\sigma_v^{\text{dSph}}=10 \; \text{km}/\text{s}$ and $\sigma_v^{\text{GC}}=150 \; \text{km}/\text{s}$ for the velocity dispersion
following~\cite{Zhao:2016xie,Boddy:2018ike}. 
We also considered the reconstruction of the radially varying velocity distribution based on the Eddington inversion method (see e.g.~\cite{Lacroix:2018qqh,Ferrer:2013cla,Biondini:2023ksj}) and found the averaged cross section to agree at the level of uncertainties typical for indirect detection constraints (see App.~\ref{app:Eddington_method}).

\begin{figure}[t]
\begin{center}
\begin{subfigure}{0.55\textwidth}
    \centering
    \includegraphics[width=\textwidth]{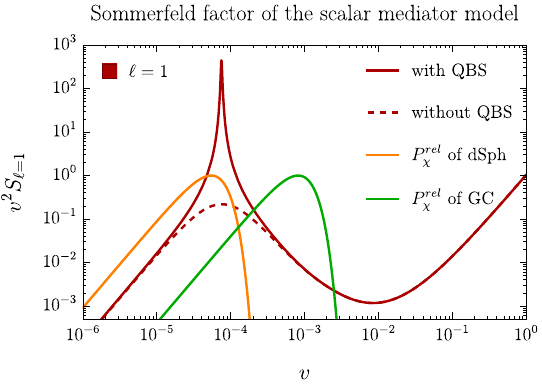}
\end{subfigure}\\
\begin{subfigure}{0.55\textwidth}
    \centering
    \includegraphics[width=\textwidth]{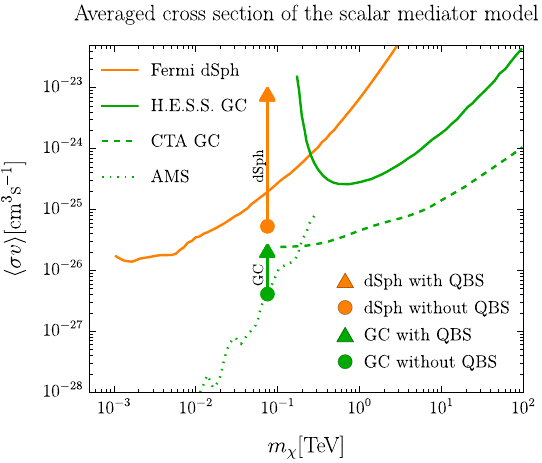}
\end{subfigure}
\end{center}
\caption{\small
Impact of QBS on the scalar mediator model. \emph{Upper panel:} Velocity-dependence of the annihilation cross section for fermionic DM annihilation into scalars for a benchmark model with and without QBS contribution. For comparison the velocity distribution within a typical dwarf spheroidal galaxy (dSph) and the Galactic Center (GC) are shown (in arbitrary units).
\emph{Lower panel:} Prediction for the velocity-averaged annihilation cross section with (triangles on top of the arrows) and without (circles at the bottom of the arrows) QBS, and for dSph (orange) as well as the GC (green) region, respectively.
Also shown are 95\%C.L. upper limits for $\bar\chi\chi\to\phi\phi\to e^+e^-e^+e^-$
derived from dSph observations by Fermi-LAT and GC data from H.E.S.S., as well as prospects for CTA. In addition, AMS-02 positron limits are displayed, see main text.}
\label{fig:scalar_results}
\end{figure}

In Fig.~\ref{fig:scalar_results} (upper panel) we show the $p$-wave Sommerfeld enhancement factor (red solid). We also include a corresponding model with $\epsilon_\phi \lesssim (\epsilon_\phi^\star)_{21}$ for comparison, for which no QBS exists (red dashed).
Furthermore, we display the assumed relative velocity distributions for the dSph (orange) and GC (green) regions. We observe that the QBS spike falls within the velocity regime probed by dSph, such that we may expect a strong impact on the corresponding limits. We find that this is indeed the case, as shown in Fig.~\ref{fig:scalar_results} (lower panel). The averaged annihilation cross section appropriate for dSph is enhanced by more than two orders of magnitude (around a factor of 150) due to the QBS spike (orange triangle versus orange circle). For the GC (green triangle and circle), the enhancement is weaker, but still amounts to a factor of around $5$.


For comparison, we also include upper limits on the averaged annihilation cross section from dSph observations by Fermi-LAT~\cite{Fermi-LAT:2015att} (orange solid) and from GC data from H.E.S.S.~\cite{HESS:2016mib} (green solid) as well as prospects for CTA~\cite{Eckner:2021jow} in the lower panel of Fig.~\ref{fig:scalar_results}. These were derived in~\cite{Profumo:2017obk,NFortes:2022dkj} for the cascade annihilation process $\bar\chi\chi\to\phi\phi\to e^+e^-e^+e^-$ with $2m_e/m_\phi=0.01$, and are thus applicable to the benchmark model, for which $2m_e/m_\phi\simeq 0.011$.
We observe that the model appears to be allowed by Fermi-LAT if the QBS spike is not taken into account, but is actually already excluded due to the QBS contribution. A similar statement holds for AMS-02 positron limits~\cite{AMS:2013fma,AMS:2021nhj}, for which however larger uncertainties related to propagation and foreground are present~\cite{Krommydas:2022loe}. As an estimate, we display AMS-02 limits on annihilation into $e^+e^-$ derived in~\cite{Bergstrom:2013jra} (green dotted).
The benchmark model without QBS has a cross section below this AMS-02 limit, such that the positron flux from cascade decay and without QBS would not be testable by AMS-02. With QBS, the positron flux could leave a detectable signature in AMS-02, which would however require a dedicated analysis for this cascade model that is beyond the scope of this work.

Overall, we find that despite the sharpness of the QBS spike, it can have a dramatic impact on indirect detection signals in $p$-wave dominated models. Furthermore, the position of the QBS spike implies that the signal strength in various astrophysical targets can be influenced very differently. This can be of phenomenological relevance when relating signals from e.g. dwarf galaxies and the GC to each other.

\subsection{Vector mediator model}
\label{sec:Vector_mediator_model}

In our second example, we consider instead a light \emph{vector} mediator $A_d^\mu$, which couples a fermionic dark matter particle $\chi$ very weakly to the SM via kinetic mixing, see e.g.~\cite{Babu:1997st,Frandsen:2011cg,Chu:2016pew,Bringmann:2016din,Cirelli:2016rnw,Baldes:2017gzu},
\begin{equation}
    \mathcal{L} = \Bar{\chi}(i \slashed{D}_d-m_\chi)\chi-\frac{1}{4}F_{d,\mu \nu}F_d^{\mu \nu}-\frac{1}{2}m_{A_d}^2 A_{d,\mu} A_d^\mu-\frac{\epsilon_{\text{mix}}}{2c_w}F_{\mu \nu}F_d^{\mu \nu}\,,
\end{equation}
where $F_d^{\mu \nu}$ is the field strength of the vector field $A_d^\mu$ and $F^{\mu \nu}$ the standard hypercharge one.
Furthermore, $D_d^\mu=\partial^\mu-ig_d A_d^\mu$ with coupling $g_d$.
The model is often referred to as dark photon or $Z^\prime$ model. Leaving aside the possibility of mass mixing, the relevant dark sector-SM portal reads 
\begin{equation}\label{eq:darkphotoncouplings}
    \mathcal{L} \supset g_f A_d^\mu \Bar{f}\gamma_\mu f\,,
\end{equation}
where $f$ stands for SM fermions and $g_f$ is related to the small kinetic mixing parameter $\epsilon_{\text{mix}}$, specifically $g_f=e\epsilon_{\text{mix}}(Y_fm_{A_d}^2/c_w^2-Q_fm_Z^2)/(m_{A_d}^2-m_Z^2)$~\cite{Cirelli:2016rnw}.  
As for the scalar model, the dark matter particles first annihilate into a pair of mediators $\Bar{\chi} \chi \to A_dA_d$, which in turn decay into SM fermions $A_d \to \Bar{f}f$. Differently from the scalar mediator case, the DM annihilation cross section into two vector mediators has both a $s$- and a $p$-wave component,\footnote{We neglect $m_{A_d}/m_\chi$ corrections and higher-order velocity corrections to the $s$-wave. Both are irrelevant in the parameter region under study.}
\begin{equation}
    \sigma v = \frac{\pi \alpha^2}{m_\chi^2} \bigg[ S_{\ell=0} +\frac{7}{12} v^2  S_{\ell=1} \bigg]\,, \label{eq:svV}
\end{equation}
where $\alpha\equiv g_d^2/(4\pi)$.
To illustrate the impact of the presence of quasi-bound states in this model, we choose a benchmark with DM mass  $m_\chi=20\,\text{TeV}$, coupling constant $\alpha=0.2410$ and, as in the scalar case, $\epsilon_\phi=0.1102$. With this choice of parameters, the obtained relic abundance is compatible with Planck observations~\cite{Planck:2018vyg}. The mediator mass is consequently fixed to be $m_{A_d}=531.1\text{GeV}$.
Here we assumed the kinetic mixing to be strong enough to keep the dark sector in kinetic equilibrium with the SM (see~\cite{Binder:2021bmg}), but weak enough to treat the mediator as effectively stable during the annihilation process, similarly as for the scalar case. The benchmark is consistent with direct detection and BBN bounds discussed in~\cite{Cirelli:2016rnw} for a wide range of $\epsilon_{\text{mix}}$.


\begin{figure}[t]
\begin{center}
\begin{subfigure}{0.55 \textwidth}
    \centering
    \includegraphics[width=\textwidth]{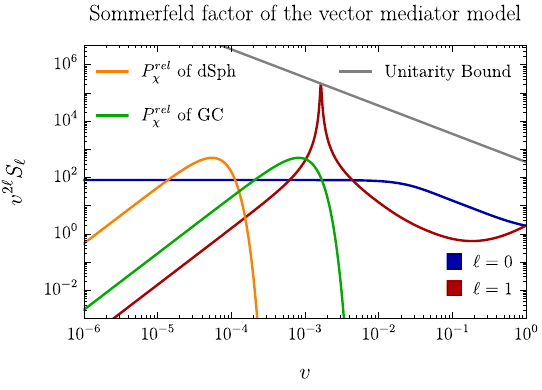}
\end{subfigure}\\
\begin{subfigure}{0.55 \textwidth}
    \centering
    \includegraphics[width=\textwidth]{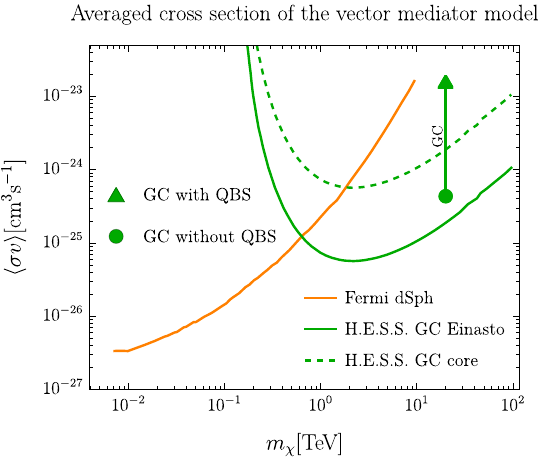}
\end{subfigure}
\end{center}
 \caption{\small
 Impact of QBS on the vector mediator model. 
 \emph{Upper panel:} $s$- (blue) and $p$-wave (red) contributions to the cross section for a vector mediator benchmark model, given by $v^{2\ell} S_\ell(v)$ for $\ell=0,1$ in Eq.~\eqref{eq:svV}. Velocity distributions in dSph (orange) and GC (green)  regions are shown for comparison as in Fig.~\ref{fig:scalar_results} (in arbitrary units), as well as the $p$-wave unitarity bound (gray).
 \emph{Lower panel:} Averaged annihilation cross section for the vector mediator benchmark model with (triangles on top of the arrows) and without (circles at the bottom of the arrows) QBS contribution, using the GC velocity distribution. In addition upper limits for cascade annihilation into light quarks as derived in~\cite{NFortes:2022dkj} are shown, for dSph Fermi-LAT and GC H.E.S.S. data, and two assumptions on the DM density profile (see~\cite{Rinchiuso:2018ajn}) for the latter.
 }
\label{fig:vector_results}
\end{figure}

 
The velocity-dependence of the annihilation cross section is shown in Fig.~\ref{fig:vector_results} (upper panel), displaying $s$- and $p$-wave contributions separately, as well as the dSph and GC velocity distributions for comparison. For the vector model, the QBS spike falls into the regime of GC velocities. Around the spike, the $p$-wave  exceeds the $s$-wave contribution by many orders of magnitude, and almost reaches the unitarity bound~\cite{Griest:1989wd}. To assess the impact on indirect detection signals, we also compute the averaged cross section following the same procedure as described in Sec.~\ref{sec:Scalar_mediator_model}. The result for the GC region is shown in Fig.~\ref{fig:vector_results} (lower panel) by the triangle (with QBS) and the circle (without QBS), respectively. Even though both $s$- and $p$-wave contribute for the vector model, the QBS enhancement of the latter still increases the total averaged cross section by a factor of approximately 35.
For comparison, we also include upper limits from Fermi-LAT dSph and H.E.S.S. GC observations derived in~\cite{Profumo:2017obk,NFortes:2022dkj} 
assuming a cascade annihilation with mediator decay into light quarks (we refer to App.~\ref{app:vectordecay} for a discussion of the applicability to the vector mediator model). H.E.S.S. limits are sensitive to the DM distribution close to the GC. The green solid line  in the lower panel of Fig.~\ref{fig:vector_results} shows the case of an Einasto profile (with parameters as in~\cite{NFortes:2022dkj}),
while the green dashed line corresponds to a cored profile (see \cite{Rinchiuso:2018ajn} and App.~\ref{app:vectordecay} for details). Thus, we find that the vector benchmark model is probed by H.E.S.S. even under conservative assumptions on the DM profile when including the QBS contribution. Without the QBS enhancement, the model would still be allowed by H.E.S.S. in this case.


\subsection{Wino}
\label{sec:wino_model}

\begin{figure}[t]
\begin{center}
\begin{subfigure}{0.47 \textwidth}
    \centering
    \includegraphics[width=\textwidth]{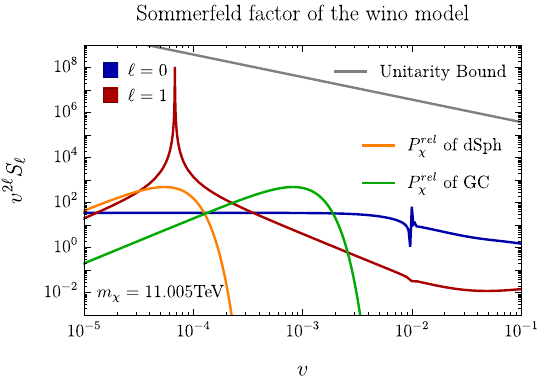}
\end{subfigure}
\begin{subfigure}{0.47\textwidth}
    \centering
    \includegraphics[width=\textwidth]{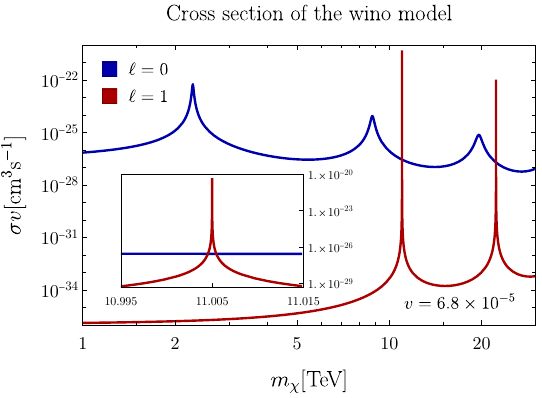}
\end{subfigure}\\
\begin{subfigure}{0.53 \textwidth}
    \centering
    \includegraphics[width=\textwidth]{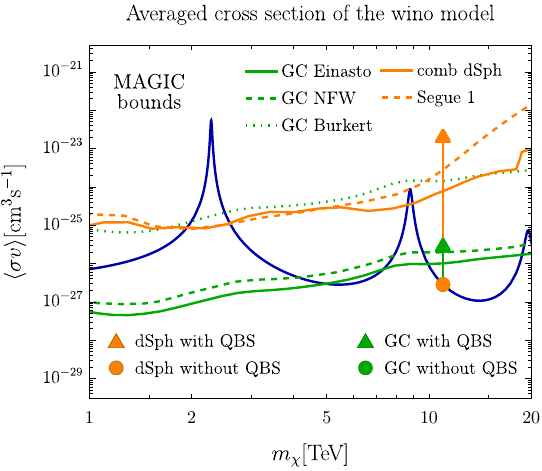}
\end{subfigure}
\end{center}
 \caption{\small Impact of QBS on wino dark matter annihilation $\chi^0\chi^0$. \emph{Upper left panel:} Velocity-dependence of the $s$- and $p$-wave Sommerfeld factors for $m_\chi=11.005$\,TeV, featuring a QBS resonance with peak at $v\simeq 6.8 \times 10^{-5}$. \emph{Upper right panel:} total $s$- and $p$-wave annihilation cross section versus wino mass for fixed relative velocity $v=6.8 \times 10^{-5}$. The inset shows a zoomed version of the first $p$-wave resonance region. \emph{Lower panel:} Velocity averaged cross section $\langle \sigma_{\gamma\gamma} v+\sigma_{\gamma Z}v/2\rangle_{s+p}$. The tip (triangles on top of the arrows) and bottom (circles at the bottom of the arrows) correspond to the sum of $s$- and $p$-wave for the wino mass $m_\chi=11.005$\,TeV with and without taking the QBS resonance into account, and for dSph (orange) and GC (green) velocity distributions, respectively. They apply to the narrow $p$-wave resonance region, while the cross section is dominated by the usual $s$-wave (blue line) for all other shown masses. We also show upper limits from MAGIC, from GC observations (green lines, for the different density profiles), from combined dSph (orange line) and from Segue 1 (dashed orange line).}
\label{fig:wino_results}
\end{figure}

We finally come back to the case of wino dark matter. As discussed in Sec.~\ref{sec:Quasi-bound_states_in_the_wino_model}, QBS are relevant for wino pair annihilation $\chi^0\chi^0$ close to $p$- and higher partial wave resonances, provided the QBS energy is smaller than the mass splitting between charged and neutral components, such that the QBS resonance occurs at a relative velocity for which the chargino pair cannot be produced on-shell. In Fig.~\ref{fig:wino_results} we show numerical results obtained using the algorithm from~\cite{Beneke:2014gja}. The first panel shows the dependence of the $\chi^0\chi^0$ annihilation cross section in the vicinity of the first $p$-wave resonance, for wino mass  $m_\chi = 11.005$\,TeV for the $s$-wave (blue) and $p$-wave (red). The latter features a QBS resonance, whose precise position depends on the QBS energy. For the mass $m_\chi=11.005$\,TeV chosen for the upper left panel of Fig.~\ref{fig:wino_results}, the location of the QBS resonance falls into the realm of typical dSph relative velocities. This corresponds to a (real part of the) QBS energy of $63.3$\,keV, 
well below the mass splitting $\delta m=164.1$\,MeV, as expected from Sec.~\ref{sec:Quasi-bound_states_in_the_wino_model}. 
In the following we assume the wino  constitutes the entirety of dark matter, independently of its production in the Early Universe.
For a wino mass above the thermal value $2.89\,$TeV~\cite{Beneke:2020vff} thermal production yields an overabundance. This
scenario therefore requires non-thermal production or a non-standard cosmological history, e.g. with a low reheating temperature or late-time
entropy injection.


The dependence of the annihilation cross section on $m_\chi$ is shown in the upper right panel of Fig.~\ref{fig:wino_results} for $v= 6.8 \times  10^{-5}$ for the $p$-wave (red) as well as the well-known $s$-wave contribution (blue). Clearly, the $p$-wave exceeds the $s$-wave in a narrow region close to the $p$-wave resonances, the first two of which can be seen in the figure. 
To resolve the resonance region, we show a zoom-in close to the first $p$-wave resonance by the inset of the upper right panel of Fig.~\ref{fig:wino_results}.

The lower panel of Fig.~\ref{fig:wino_results} displays the averaged cross section $\langle \sigma_{\gamma\gamma} v+\sigma_{\gamma Z}v/2\rangle$, including $s$- and $p$-wave contributions, and highlights the QBS effects for the wino model with mass $m_\chi=11.005$\,TeV (triangles at the tip of the orange and green arrows for the dSph and GC velocity distributions, respectively).  
We also display the result when omitting the QBS (circles at the bottom of the arrows). We note that even after performing a velocity average, the cross section is enhanced by a factor around $7500$ ($10$) for the dSph (GC) regions due to the QBS resonance. Away from the rather narrow $p$-wave resonance regions, the cross section is dominated by the usual $s$-wave contribution (blue line). We compare to upper limits from MAGIC observations of the GC region~\cite{MAGIC:2022acl} for various assumptions on the dark matter profile, and to MAGIC limits obtained from observing the dwarf galaxy Segue 1~\cite{Aleksic:2013xea,MAGIC:2021mog} as well as a combined limit from Coma Berenices, Ursa Major II, Draco and Segue 1~\cite{MAGIC:2021mog}. We note uncertainties related to 1) the dark matter content of the dwarf galaxies, and 2) due to velocity-average given the peculiar velocity dependence of the QBS resonance. Considering $J$-factors for Segue 1 given in~\cite{Aleksic:2013xea,MAGIC:2021mog,Boddy:2017vpe} the former can be estimated to be around  a factor of three. For the latter we compare the velocity average using a Maxwellian with fiducial velocity dispersion as explained in Sec.~\ref{sec:Scalar_mediator_model} to those obtained from the Eddington inversion method, finding uncertainties at the level of $30\%$ for Segue 1 and up to a factor two for all considered dwarf galaxies (see App.~\ref{app:Eddington_method}).

Notably, we find that existing dwarf galaxy observations by MAGIC~\cite{MAGIC:2021mog} are actually already sensitive to  wino dark matter with mass $m_\chi=11.005$\,TeV  due to the QBS resonance. Indeed, the averaged cross section exceeds the nominal 95\% C.L. upper limit from Segue 1 alone by a factor $15$, and the combined dwarf limit by a factor $55$, which is significant even when considering the uncertainties discussed above. 
When neglecting the QBS, this wino mass would not be currently probed by neither GC nor dwarf gamma-ray data. The mass range that can be probed is however extremely narrow, and can be bracketed by the masses for which the QBS appears and for which its velocity peak is shifted to above the characteristic DM velocity  $v\simeq 2\cdot 10^{-4}$ of the dSph galaxy, giving a mass range of about 1\,GeV width. We thus conclude that the wino with this particular mass can already be tested by virtue of the QBS feature. Moreover, it can be  probed with dwarf galaxies rather than with GC observations, due to the particular enhancement of the cross section at relative velocities matching those inside typical dwarf galaxies.


\section{Conclusions}
\label{sec:Conclusion}

In this work we provided a detailed quantum-mechanical understanding of certain peculiar features in the Sommerfeld enhancement for $p$- and higher partial waves. The most prominent feature are  spikes in the velocity-dependence of the Sommerfeld-enhanced annihilation cross section, that occur at particular values of the relative velocity of the incoming pair of dark matter particles, for model parameters in the vicinity of $p$- or higher $\ell$-wave Sommerfeld resonances.

We find that these spikes can be explained by the existence of  quasi-bound states, i.e. metastable states with positive energy, that are sustained by the interplay of an attractive Yukawa force and the centrifugal barrier. The spike occurs when the kinetic energy of the relative motion of the incoming dark matter pair matches the quasi-bound state energy. We provide a detailed understanding of this phenomenon, both on a qualitative level as well as quantitatively via full numerical results. We also offer analytical results based on the WKB approximation that capture the main qualitative features, as well as useful quantitative estimates of the QBS width and residue.

The existence of QBS with angular momentum $\ell\geq 1$ is closely linked to Sommerfeld resonances in the $\ell$th partial wave. As is well-known, the latter are related to model parameters (typically dark matter or mediator masses) for which the Yukawa-like attractive potential leads to zero-energy bound states with angular momentum $\ell$. For slightly different mass values, the zero-energy bound state can either become a true, weakly bound state, or, for $\ell\geq 1$, alternatively a quasi-bound state with small positive energy. In the latter case, a spike feature emerges in the form of a Breit-Wigner resonance for the relative velocity matching the QBS energy. This property can therefore be seen as a dynamical generation of a ``super-resonant'' enhancement~\cite{Beneke:2022rjv}.

Our results also explain another peculiar feature: the universal velocity-scaling of the Sommerfeld factors for all $\ell\geq 1$ at the precise location of the Sommerfeld resonance, $S_\ell\propto v^{-4}$, as opposed to the well-known $S_0\propto v^{-2}$ for the $s$-wave. This property is related to the velocity-scaling of the QBS width for $\ell\geq 1$.\footnote{These observations  emphasize that the commonly used Cassel approximation~\cite{Cassel:2009wt} for the Sommerfeld factor based on analytic results in a modified Hulth\'en potential should not be used for $\ell\geq 1$, since it misses both the QBS spikes and the universal velocity scaling properties.  Based on numerical comparisons, the failure of the approximation close to Sommerfeld resonances has in fact already been remarked upon in the original work.}

The results presented in this work have several implications. First, close to a QBS resonance the $p$-wave contribution to the annihilation cross section can exceed the $s$-wave one  by several orders of magnitude, even after averaging over a Maxwellian velocity distribution. We showed that this modifies conclusions drawn from  indirect detection bounds for the viability of models with light scalar or vector mediators. We also considered wino dark matter, finding a sizable QBS enhancement for wino masses around $11$\,TeV, related to the location of the first $p$-wave Sommerfeld resonance. We expect that QBS are also important for dark matter residing in other electroweak multiplets~\cite{Bottaro:2021snn, Baumgart:2023pwn}. Notably, since the QBS spike has a sharp dependence on relative velocity, its position can affect different targets for indirect detection searches such as dwarf spheroidal galaxies or the Galactic Center region very differently. This is of direct phenomenological relevance when relating signals or constraints from dwarfs and the Milky Way Center to each other. Indeed, we find that (non-thermally produced) wino dark matter in a very narrow mass range around $11\,$TeV can already be tested with existing MAGIC dwarf galaxy observations, while being compatible with GC constraints.  Second, we note that the quasi-bound state resonance is also relevant  for other processes involving a non-relativistic pair of dark matter particles subject to Yukawa-like forces, specifically bound-state formation and self-scattering, and explains spikes in their velocity dependence for $\ell\geq 1$ partial waves~\cite{Ding:2021zzg,Kamada:2023iol,Biondini:2023ksj,Petraki:2016cnz,Tulin:2012wi}.

\subsubsection*{Acknowledgments}
We thank Ayuki Kamada and Stefan Lederer for discussion. We acknowledge support by the DFG Collaborative Research Institution Neutrinos and Dark Matter in Astro- and Particle Physics (SFB 1258) and the Excellence Cluster ORIGINS - EXC-2094 - 390783311.


\appendix

\section{WKB details}
\label{app:WKB_details}

In this appendix we provide a derivation of the expression \eqref{S_WKB} for the WKB approximation of the Sommerfeld factor in presence of a quasi-bound state.
Since we consider a low-velocity scattering state, we assume that the energy $\epsilon_v^2$ is lower than the height of the barrier. Therefore we obtain three classical turning points $x_i$, defined by the solution of the equation 
\begin{equation}
    \label{TP_def}
    \epsilon_v^2=V_{\ell}^{\text{eff}}(x_{i}).
\end{equation}
At these points the reduced wave function $u_{\ell}(x)$ has to be matched by means of the standard WKB procedure.
The result of this procedure reads  \cite{Merzbacher:1970,Scrucca:QP3}
\begin{equation}
    \label{eq:WKBwf}
    u_{\ell}(x) \approx
        \begin{cases}
            \frac{A_\ell}{2\sqrt{p(x)}} e^{-\int_{x}^{x_1} dx' \, p(x')} & x<x_1\\[0.1cm]
            \frac{A_{\ell}}{\sqrt{p(x)}} \sin\left(\int_{x_1}^x dx' \, p(x')+\frac{\pi}{4}\right) & x_1<x<x_2 \\
            \frac{A_{\ell}}{\sqrt{p(x)}} \bigg[\frac{1}{2} s_w e^{-\int_{x_2}^x dx' \, p(x')} + c_w e^{\int_{x_2}^x dx' \, p(x')}\bigg] & x_2<x<x_3 \\[0.3cm]
            \frac{A_{\ell}}{\sqrt{p(x)}} \bigg[\left(c_w e^{I_b}+\frac{i}{4}s_we^{-I_b}\right) e^{i\int_{x_3}^x dx' \, p(x')-\frac{i\pi}{4}} & \\ \hspace*{0.9cm}+ \left(c_w e^{I_b}-\frac{i}{4}s_w e^{-I_b}\right) e^{-i\int_{x_3}^x dx' \, p(x')+\frac{i\pi}{4}}\bigg] & x_3<x\,,
        \end{cases}
\end{equation}
where we defined $c_w \equiv \cos(I_w)$, $s_w \equiv \sin(I_w)$,
\begin{equation}
    \label{eq:exactQuantization}
    I_w \equiv \int_{x_1}^{x_2} dx' \, p(x') \quad \text{and} \quad
    I_b \equiv \int_{x_2}^{x_3} dx' \, p(x')\,.
\end{equation}
First, we obtain the Sommerfeld factor by means of \eqref{eq:SEl}, together with the asymptotic condition $u_{\ell}(x) \to \sin(\epsilon_v x - \frac{\ell \pi}{2}+\delta_\ell)$
as $x\to\infty$ of the scattering state. This fixes the normalization constant to
\begin{equation}
    A_{\ell}=\frac{\sqrt{\epsilon_v}}{2\left|c_w e^{I_b}+\frac{i}{4}s_w e^{-I_b}\right|}\,,
\end{equation}
hence the wave function in the region $x<x_1$ reads
\begin{equation}
    \label{eq:WKBwf01}
    u_{\ell}(x)=\frac{\sqrt{\epsilon_v} \exp\left\{-\int_{x}^{x_1} dx' \, p(x')\right\}}{4\left|c_w e^{I_b}+\frac{i}{4}s_w e^{-I_b}\right| \sqrt{p(x)}}\,.
\end{equation}
We aim to compute the $(\ell+1)$-th derivative of $u_{\ell}$, therefore we expand for $x \to 0$ and exploit
\begin{equation}
    \sqrt{p(x)}=\sqrt{\frac{\ell+1/2}{x}}+\mathcal{O}(\sqrt{x})\,.
\end{equation}
The integral $\int_x^{x_1} dx' \, p(x')$ in \eqref{eq:WKBwf01} is logarithmically divergent for $x \to 0$ due to the centrifugal term $V_{\ell}^{\text{eff}} \sim \frac{(\ell+1/2)^2}{x^2}$, hence $\exp\left\{-\int_x^{x_1} dx' \, p(x')\right\} =a x^{\ell+1/2}+\mathcal{O}(x^{\ell+3/2})$. Therefore, in order to extract exactly the coefficient $a$, we insert $1=\frac{1}{x} x$ and integrate by parts. This results in
\begin{equation}
    \label{eq:byparts}
    -\int_x^{x_1} dx' \frac{1}{x'} x'p(x')= \ln(x) x p(x)
    + \int_x^{x_1} dx' \, \ln(x') \frac{d}{dx'}[x' p(x')]\,,
\end{equation}
where of the boundary terms vanishes because $p(x_1)=0$. In the limit $x \to 0$, $x p(x) \to (\ell+1/2)+\mathcal{O}(x)$ and the integral on the right-hand side of \eqref{eq:byparts} is finite, hence:
\begin{equation}
    -\int_x^{x_1} dx' p(x') = (\ell+1/2)\ln(x)
    +\int_0^{x_1} dx' \, \ln(x') \frac{d}{dx'}[x'p(x')]+\mathcal{O}(x)\,.
\end{equation}
It follows that \eqref{eq:WKBwf01} can be expanded as
\begin{equation}
    u_{\ell}(x) = \frac{\sqrt{\epsilon_v} x^{\ell+1}}{4\sqrt{\ell+1/2}\left|c_w e^{I_b}+\frac{i}{4}s_w e^{-I_b}\right|}
    \times \exp \left\{\int_0^{x_1} dx' \, \ln(x') \frac{d}{dx'}[x'p(x')]\right\}+ \mathcal{O}(x^{\ell+2})\,.
\end{equation}
Defining the exponential factor 
\begin{equation}
\sqrt{\mathcal{P}} = \exp \left\{\int_0^{x_1} dx' \, \ln(x') \frac{d}{dx'}[x'p(x')]\right\},
\end{equation}
and taking ($\ell+1$)-th derivative evaluated at the origin, we find
\begin{equation}
    \label{eq:WKBwfDeriv}
    \partial_x^{\ell+1}u_\ell(0) = \frac{(\ell+1)! \,\sqrt{\mathcal{P}}\,\sqrt{\epsilon_v}}{4\sqrt{\ell+1/2}\left|c_w e^{I_b}+\frac{i}{4}s_w e^{-I_b}\right|}\,.
\end{equation}
By substituting \eqref{eq:WKBwfDeriv} into \eqref{eq:SEl}, we obtain
\begin{equation}
    \label{S_WKB_app1}
    S_{\ell} = \frac{[(2\ell+1)!!]^2\,\mathcal{P}}{16 (\ell+1/2) \epsilon_v^{2\ell+1} \left|c_w e^{I_b}+\frac{i}{4}s_w e^{-I_b}\right|^2}\,.
\end{equation}
It is convenient to rewrite the above expression in terms of the semi-classical width $\gamma$, which is defined as
\begin{equation}
    \label{width}
    \gamma \equiv \frac{e^{-2I_b}}{T},
\end{equation}
where $T$ is the classical oscillation period in the well
\begin{equation}
    \label{period}
    T \equiv 2\frac{dI_w}{d\epsilon_v^2}.
\end{equation}
In terms of $\gamma$ and $T$, expression \eqref{S_WKB_app1} can be rewritten as:
\begin{equation}
    \label{S_WKB_app2}
    S_{\ell} = \frac{[(2\ell+1)!!]^2\,\mathcal{P}}{(2\ell+1) \epsilon_v^{2\ell+1} T |s_w|^2}\times \frac{\gamma/2}{\left( \frac{2 \cot{I_w}}{T}\right)^2+ \left(\frac{\gamma}{2}\right)^2}\,,
\end{equation}
which is the WKB approximation of the Sommerfeld factor for a {\em generic} low-velocity scattering state in a potential with three classical turning points.

We remark on the velocity dependence of the period $T$ and of the width $\gamma$ in the limit $\epsilon_v \ll 1$. From \eqref{period}, the explicit expression of $T$ reads 
\begin{equation}
    T(\epsilon_v)=\int_{x_1(\epsilon_v)}^{x_2(\epsilon_v)} \frac{dx'}{\sqrt{\epsilon_v^2-V_{\ell}^{\text{eff}}(x')}} \,,
\end{equation}
where $x_{1,2}(\epsilon_v)$ are the classical turning points defined in \eqref{TP_def}.
For $\epsilon_v^2>\min_x\{V_\ell^{\text{eff}}(x)\}$, we have $\epsilon_v^2>V_\ell^{\text{eff}}(x)$ for $x_1(\epsilon_v) < x < x_2(\epsilon_v)$, therefore the period is strictly positive. This is also true for $\epsilon_v=0$ because $0>\min_x \{V_\ell^{\text{eff}}(x)\}$. The integrand is singular at the integration boundaries $x_i(\epsilon_v)$, close to which it behaves as 
\begin{equation}
    \frac{1}{\sqrt{\epsilon_v^2-V_\ell^{\text{eff}}(x)}} \to \frac{1}{\sqrt{-(x-x_i(\epsilon_v))V_\ell^{\prime\,\text{eff}}(x_i(\epsilon_v))}}.
\end{equation}
where the derivative of the effective potential does not vanish at the turning points. The same argument applies to the  $\epsilon_v=0$ case, where the classical turning points are
\begin{equation}
    \label{TP_0}
    x_1(0)=-\frac{1}{\epsilon_\phi}W_0(-(\ell+1/2)^2 \epsilon_\phi) \quad \text{and} \quad x_2(0)=-\frac{1}{\epsilon_\phi} W_{-1}(-(\ell+1/2)^2\epsilon_\phi),
\end{equation}
where $W_{0,-1}$ are branch cuts of the Lambert W function and the derivative of the potential reads
\begin{equation}
    V_{\ell}^{\prime\,\text{eff}}(x_{1,2}(0))=\frac{(\ell+1/2)^2}{[x_{1,2}(0)]^3}(\epsilon_\phi x_{1,2}(0)-1).
\end{equation}
Therefore the period $T(\epsilon_v)$ approaches a positive finite value $0<T(0)<\infty$. The  explicit expression for the width $\gamma$ reads
\begin{equation}
    \gamma(\epsilon_v)=\frac{1}{T(\epsilon_v)}\exp\left\{-2\int_{x_2(\epsilon_v)}^{x_3(\epsilon_v)} dx' \, \sqrt{V_\ell^{\text{eff}}(x')-\epsilon_v^2}\right\}.
\end{equation}
Since the period approaches the constant value $T(0)$, we focus on the velocity dependence of the exponential 
\begin{equation}
    \label{width_exp}
    \exp{-2\int_{x_2(\epsilon_v)}^{x_3(\epsilon_v)} dx' \sqrt{\frac{(\ell+1/2)^2}{x'^2}-\frac{e^{-\epsilon_\phi x'}}{x'}-\epsilon_v^2}}\,.
\end{equation}
As $\epsilon_v \ll 1$, the lower integration limit $x_2$ approaches the constant value \eqref{TP_0}, while the upper one, $x_3$, determined by \eqref{TP_def}, approaches $\infty$ as 
\begin{equation}
    x_3=\frac{\ell+1/2}{\epsilon_v}+\mathcal{O}\left(e^{-\frac{\epsilon_\phi(\ell+1/2)}{\epsilon_v}}/\epsilon_v^2\right).
\end{equation}
Therefore, we take the limits $\epsilon_v \to 0$ and $x \to \infty$, together with the condition $x \epsilon_v \lesssim \ell+1/2$. Then, the integrand can be expanded as
\begin{equation}
    \sqrt{\frac{(\ell+1/2)^2}{x^2}-\epsilon_v^2-\frac{e^{-\epsilon_\phi x}}{x}}=\sqrt{\frac{(\ell+1/2)^2}{x^2}-\epsilon_v^2}+\mathcal{O}\left(\frac{e^{-\epsilon_\phi x}}{x\sqrt{\frac{(\ell+1/2)^2}{x^2}-\epsilon_v^2}}\right),
\end{equation}
where we emphasize that the neglected part leads to a finite integrand due to the exponential suppression of the factor $e^{-\epsilon_\phi x}$.
Finally, the width can be written as
\begin{eqnarray}
    \gamma(\epsilon_v)&=&\frac{1}{T(0)}\exp\left\{-2\int_{x_2(0)}^{\frac{\ell+1/2}{\epsilon_v}} dx'\, \sqrt{\frac{(\ell+1/2)^2}{x'^2}-\epsilon_v^2}+\mathcal{O}(1)\right\}
    \nonumber\\
    &=&\frac{1}{T(0)}\exp\big[(2\ell+1)\ln(\epsilon_v)+\mathcal{O}(1)\,\big] \sim \epsilon_v^{2\ell+1}.
\end{eqnarray}
We remark that $\gamma \sim \epsilon_v^{2\ell+1} \ll \epsilon_v^2 \ll 1$ implies that the width is smaller than the energy for $\ell \geq 1$.

Next, we require the existence of a positive-energy metastable state (the quasi-bound state), that is, we ask for the existence of a solution with wave function that behaves like a progressive wave $e^{i \epsilon x}$ for $x \to \infty$. By applying this boundary condition to \eqref{eq:WKBwf}, we find that for $x>x_3$ the coefficient $\left(c_w e^{I_b}-\frac{i}{4}s_w e^{-I_b}\right)$ of the regressive wave component $e^{-i \int_{x_3}^x dx' \, p(x')}$  must vanish, which implies
\begin{equation}
    \label{complex_quantization}
    \cot(I_w) =\frac{i}{4} e^{-2I_b}\,.
\end{equation}
Since the right-hand side is complex, it implies that also the energy eigenvalues that solve this equation need to be complex and we write them as $\epsilon^2=\epsilon_{n\ell}^2-i\frac{\gamma_{n\ell}}{2}$. We now assume that the imaginary part of the energy is much smaller than the real part, i.e. $\gamma_{n\ell} \ll \epsilon_{n\ell}^2$. Therefore, we expand the left- and right-hand sides of \eqref{complex_quantization} for $\gamma_{n\ell} \ll \epsilon_{n\ell}^2$, leading to
\begin{equation}
    \left.\cot(I_w)\right|_{\epsilon^2=\epsilon_{n\ell}^2}-i \frac{\gamma_{n\ell}}{2}\left.\frac{d \cot(I_w)}{d\epsilon^2}\right|_{\epsilon^2=\epsilon_{n\ell}^2}=\frac{i}{4}\left.e^{-2I_b}\right|_{\epsilon^2=\epsilon_{n\ell}^2}.
\end{equation}
The above equation can be separated into its real and imaginary components
\begin{equation}
    \left.\cot(I_w)\right|_{\epsilon^2=\epsilon_{n\ell}^2}=0 \quad \text{and} \quad \gamma_{n\ell}\left.\frac{d \cot(I_w)}{d \epsilon^2}\right|_{\epsilon^2=\epsilon_{n\ell}^2}=-\left.\frac{e^{-2I_b}}{2}\right|_{\epsilon^2=\epsilon_{n\ell}^2},
\end{equation}
which leads to a quantization condition for the real part, $\epsilon_{n\ell}$, of the energy eigenvalue, 
\begin{equation}
    \label{real_quantization}
    \left.I_w\right|_{\epsilon^2=\epsilon_{n\ell}^2}=\left(n_r+\frac{1}{2}\right)\pi\,,
\end{equation}
where $n_r$ is the radial quantum number and $n=n_r+\ell$. Regarding the imaginary component, it leads to the expression for the width $\gamma_{n\ell}=\gamma(\epsilon_{n\ell})$, consistently with the definition \eqref{width}.

With these results at hand, we return to the expression \eqref{S_WKB_app2} for the Sommerfeld factor and we focus on a scattering state with energy $\epsilon_v^2$ close to the quasi-bound state energy $\epsilon_{n\ell}^2$. By expanding \eqref{S_WKB_app2} for $\epsilon_v^2 \to \epsilon_{n\ell}^2$, we exploit $\gamma \to \gamma_{n\ell}$, $|s_w|^2 \to 1$, $T \to T_{n\ell}$ and $\mathcal{P} \to \mathcal{P}_{n\ell}$. The only non-trivial limit comes from the $\cot(I_w)$ in the denominator, which reads
\begin{eqnarray}
    \cot(I_w)&=&\left.\cot(I_w)\right|_{\epsilon^2=\epsilon_{n\ell}^2} + (\epsilon_v^2-\epsilon_{n\ell}^2)\left.\frac{d \cot(I_w)}{d\epsilon^2}\right|_{\epsilon^2=\epsilon_{n\ell}^2}+\mathcal{O}\left((\epsilon_v^2-\epsilon_{n\ell}^2)^2\right) \nonumber \\
    &=&-(\epsilon_v^2- \epsilon_{n\ell}^2)\frac{T_{n\ell}}{2}+\mathcal{O}\left((\epsilon_v^2-\epsilon_{n\ell}^2)^2\right).
\end{eqnarray}
By substituting the above limits into \eqref{S_WKB_app2}, we finally obtain:
\begin{equation}
    S_\ell=\frac{[(2\ell+1)!!]^2\,\mathcal{P}_{n\ell}}{(2\ell+1)\epsilon_{n\ell}^{2\ell+1}T_{n\ell}} \times \frac{\gamma_{n\ell}/2}{(\epsilon_v^2-\epsilon_{n\ell}^2)^2+(\gamma_{n\ell}/2)^2} \,.
\end{equation}


\section{Eddington method}
\label{app:Eddington_method}

In Sec.~\ref{sec:Examples} we computed the  averaged annihilation cross section assuming a Maxwellian velocity distribution with fixed velocity dispersion for dSph and GC regions. In order to test this assumption, we compare to an average computed with a velocity distribution obtained via the Eddington inversion method~\cite{10.1093/mnras/76.7.572,GalacticDynamics,Lacroix:2018qqh}. This method exploits that for stationary and (in the simplest case) spherically symmetric density distributions, the complete phase-space distribution can be reconstructed when assuming ergodicity.

In our analysis we follow~\cite{Ferrer:2013cla}. As a benchmark for dSph we assume a Navarro–Frenk–White density profile with parameters adapted to the Milky Way satellite dwarf galaxy Segue 1, while for the GC we assume an Einasto density profile and add baryonic contributions from the galactic bulge and disk to the gravitational potential, see~\cite{Ferrer:2013cla} for details.
Once the DM density profile and the gravitational potential has been chosen, the velocity distribution $P_\chi^{rel}(r,v)$ is fixed by the Eddington inversion formula~\cite{GalacticDynamics}. Since the velocity distribution now depends on the radial coordinate, it is not possible to factorize the astrophysical contribution (the $J$-factor), from the particle physics contribution, i.e. the  averaged annihilation cross section.
Therefore, it is convenient to define an \emph{effective} $J$-factor~\cite{Boddy:2017vpe,Boddy:2018ike,Boddy:2019qak}, which contains the velocity dependent part of the annihilation cross section,
\begin{equation}
    J_{\ell} \equiv \int_{\Delta \Omega} d\Omega\int ds \, \rho_\chi^2(r) \int dv \, P_\chi^{rel}(r, v) v^{2\ell} S_{\ell}(v)\,,
\end{equation}
where $\Delta \Omega$ is the region of interest (ROI), $ds$ is the integral over the line-of-sight, and $\rho_\chi$ is the DM density profile. The ROI for the GC has been chosen consistently with~\cite{Profumo:2017obk,NFortes:2022dkj}.


\begin{table}[t]
\centering
\begin{tabular}{|c|c|c|}
\hline
$J_{\ell=1}$ & Fixed $[\text{GeV}^2\text{cm}^{-5}]$ & Eddington $[\text{GeV}^2\text{cm}^{-5}]$ \\ \hline
scalar dSph & $1.32 \times 10^{21}$ & $1.15 \times 10^{21}$ \\ \hline
scalar GC & $4.94 \times 10^{20}$ & $2.93 \times 10^{20}$ \\ \hline
vector dSph & $3.56 \times 10^{19}$ & $4.66 \times 10^{19}$ \\ \hline
vector GC & $3.46 \times 10^{25}$ & $7.41 \times 10^{25}$ \\ \hline
wino dSph & $2.26 \times 10^{25}$ & $1.81 \times 10^{25}$ \\ \hline
wino GC & $4.49 \times 10^{24}$ & $2.53 \times 10^{24}$ \\ \hline
\end{tabular}
\caption{\small We report the effective $J$-factor values obtained with a fixed Maxwellian velocity distribution. We compare them with the results obtained with a radially dependent velocity distribution by means of the Eddington inversion formula. We give results for the scalar mediator, vector mediator and wino benchmark models as well as for dSph and GC regions, respectively. As a benchmark for the Eddington method applied to dSph, we choose Segue~1.}
\label{tab:Eddington}
\end{table}


We compare the effective $J$-factor for a fixed Maxwellian velocity distribution as in Sec.~\ref{sec:Examples} with the one for the velocity distribution obtained by means of the Eddington inversion formula, leading to the results reported in Tab.~\ref{tab:Eddington} for the effective $p$-wave $J$-factors of the scalar, the vector benchmark model and the wino with QBS. The difference between the two approaches amounts to at most approximately a factor two for the considered scenarios, and to about $30\%$ for the wino dSph case. For the wino, we checked that the differences are below a factor two also for the other dwarf galaxies (Coma Berenices, Ursa Major II, Draco) considered in~\cite{MAGIC:2021mog}.


\section{Vector mediator details}
\label{app:vectordecay}

In this appendix we provide some details on the cascade annihilation spectrum of the vector mediator model discussed in Sec.~\ref{sec:Vector_mediator_model}, and the impact of a cored profile on H.E.S.S. GC limits, respectively.

\begin{figure}[t]
    \centering
    \includegraphics[width=0.6\textwidth]{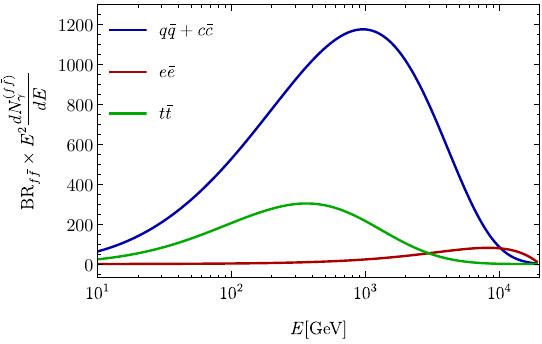}
\caption{\small Gamma-ray spectra from cascade annihilation $\chi\bar\chi\to A_dA_d$ and $A_d\to f\bar f$ for the vector mediator benchmark model with $m_\chi=20\,\text{TeV}$ and $m_{A_d}=531\,\text{GeV}$, for the relevant decay channels $f\bar f$ as shown in the legend. }
\label{fig:spectra}
\end{figure}

The gamma-ray spectrum produced in the annihilation $\bar\chi\chi\to A_dA_d$ and subsequent decay $A_d\to f\bar f$ can be computed by boosting the gamma-ray spectrum $dN^0_\gamma/dE$ for $A_d$ decaying at rest via~\cite{Escudero:2017yia}
\begin{equation}
    \frac{dN_\gamma}{dE}=2 \frac{m_{A_d}}{m_\chi} \int_{E_{min}}^{E_{max}} \frac{dE^0}{E^0} \frac{dN_\gamma^0}{dE^0}\,,
\end{equation}
where $E_{min} \equiv m_{A_d} E/(2m_\chi)$ and $E_{max}=\min\left(2m_\chi E/m_{A_d},m_{A_d}/2\right)$. The spectra $dN^0_\gamma/dE$ further depend on the branching fractions of the decay $A_d\to f\bar f$, that are fixed by the couplings generated via kinetic mixing, see~\eqref{eq:darkphotoncouplings} and~\cite{Cirelli:2016rnw}.
For the vector mediator benchmark model from Sec.~\ref{sec:Vector_mediator_model} with $m_{A_d}\simeq 530 \text{GeV}$, the dominant $A_d$ decay channels are: $23\%$ $q\bar q\equiv u\bar u,d\bar d,s\bar s$, $15\%$ $c\bar c$, $13\%$ $e^+e^-$, $11\%$ $t\bar t$ and more minor contributions. Using the gamma-ray spectra  for the respective channels provided in~\cite{Cirelli:2010xx}, we find the boosted gamma-ray spectrum for the benchmark model with $m_\chi=20\,$TeV shown in Fig.~\ref{fig:spectra}.


We note that in the limit $m_f\ll m_{A_d}\ll m_\chi$, the mapping from $dN^0_\gamma/dE$ to $dN_\gamma/dE$ becomes approximately independent of $m_{A_d}$. Furthermore, the gamma-ray spectrum can be approximated by a spectrum for which only decay into light quarks is assumed within a factor less than two. This justifies to compare the annihilation cross section of the benchmark model to upper limits on cascade annihilation  provided in \cite{NFortes:2022dkj} (see Sec.~\ref{sec:Vector_mediator_model}), that were obtained under the assumptions of decay into light quarks and for $m_f\ll m_{A_d}\ll m_\chi$.

The H.E.S.S. bounds shown in Fig.~\ref{fig:vector_results} for the cored density profile have been obtained as follows: The profile with core radius $r_c$ is parameterized by $\rho(r)=\rho_{Ein}(r)$ for $r>r_c$ and $\rho(r)=\rho_{Ein}(r_c)$ for $r<r_c$, where $\rho_{Ein}(r)$ is the fiducial Einasto profile for the GC~\cite{Ferrer:2013cla}.
Disregarding changes in the velocity distribution (see App.~\ref{app:Eddington_method}), gamma-ray limits weaken by a factor given by the ratio of $J$-factors obtained for the cored and Einasto profiles. For a core radius of $1\,$kpc, this amounts to approximately a factor $10$~\cite{Rinchiuso:2018ajn}.

\bibliography{main.bib}
\end{document}